\documentstyle[12pt,subeqnarray,cite,epsf]{article}

\newlength{\titlesep}
\newlength{\authorsep}
\makeatletter

\renewcommand{\thesection}{\Roman{section}}

\def\fnum@figure{FIG.~\thefigure}

\newcommand{\reffig}[1]{Fig.~\protect\ref{#1}}

\@addtoreset{equation}{section}
\renewcommand{\theequation}{\arabic{section}.\arabic{equation}}
\newcommand{\refeq}[1]{Eq.~(\protect\ref{#1})}

\renewcommand{\abstract}{\if@twocolumn
  \section*{Abstract}
  \else
  \begin{center}
    {\bf Abstract\vspace{-.5em}\vspace{0pt}}
  \end{center}
  \fi}
\renewcommand{\endabstract}{\if@twocolumn\else\endquotation\fi}

\renewcommand{\appendix}{\par
  \setcounter{section}{0}
  \setcounter{subsection}{0}
  \renewcommand{\thesection}{Appendix~\Alph{section}}
  \renewcommand{\theequation}{(\Alph{section}.\arabic{equation})}}

\newcommand{\thismonth}{\ifcase\month\or
 January\or February\or March\or April\or May\or June\or
 July\or August\or September\or October\or November\or December\fi
 \space \number\year}

\newcommand{\preprintnumber}[1]
{\begin{flushright}
  \begin{tabular}{l}#1\end{tabular}
  \end{flushright}}

\makeatother

\newcommand{\ie}{{\it i.e.\/}}
\newcommand{\etal}{{\it et al.\/}}
\newcommand{\etc}{{\it etc.\/}}
\newcommand{\ol}[1]{\overline{#1}}
\newcommand{\wt}[1]{\widetilde{#1}}
\newcommand{\vev}[1]{\left\langle #1 \right\rangle}
\newcommand{\hc}{{\rm h.\,c.}\,}
\newcommand{\sign}{\mathop{\rm sign}\nolimits}
\newcommand{\br}{\mathop{\rm B}\nolimits}   
\newcommand{\im}{\mathop{\rm Im}\nolimits} 
\newcommand{\gsim}%
{\mathrel{\mbox{\raisebox{-1.0ex}
    {$\stackrel{\displaystyle >}{\displaystyle \sim}$}}}}
\newcommand{\lsim}%
{\mathrel{\mbox{\raisebox{-1.0ex}
    {$\stackrel{\displaystyle <}{\displaystyle \sim}$}}}}

\newcommand{\bb}{$B^0$--$\ol{B}^0$}
\newcommand{\kk}{$K^0$--$\ol{K}^0$}
\newcommand{\ek}{\mbox{$\epsilon_K$}}
\newcommand{\dmb}{\mbox{$\Delta m_B$}}
\newcommand{\dmbs}{\mbox{$\Delta m_{B_s}$}}
\newcommand{\bsg}{\mbox{$b\rightarrow s\, \gamma$}}
\newcommand{\bsll}{\mbox{$b\rightarrow s\, l^+\, l^-$}}
\newcommand{\bsmm}{\mbox{$b\rightarrow s\, \mu^+\, \mu^-$}}
\newcommand{\bsnn}{\mbox{$b\rightarrow s\, \nu\, \ol{\nu}$}}

\newcommand{\kpnn}{\mbox{$K\rightarrow \pi\, \nu\, \ol{\nu}$}}
\newcommand{\kppnn}{\mbox{$K^+\rightarrow \pi^+\, \nu\, \ol{\nu}$}}
\newcommand{\klpnn}{\mbox{$K_L\rightarrow \pi^0\, \nu\, \ol{\nu}$}}
\newcommand{\kppen}{\mbox{$K^+\rightarrow \pi^0\, e^+\, \nu_e$}}

\newcommand{\Journal}[4]{{#1} {\bf #2}, {#4} {(#3)}}
\newcommand{\xxx}[1]{{\tt #1}}
\newcommand{\pl}{\sl Phys.~Lett.}
\newcommand{\plb}{\sl Phys.~Lett.~{\bf B}}
\newcommand{\prp}{\sl Phys.~Rep.}

\newcommand{\prd}{\sl Phys.~Rev.~{\bf D}}
\newcommand{\prl}{\sl Phys.~Rev.~Lett.}
\newcommand{\np}{\sl Nucl.~Phys.}
\newcommand{\npb}{\sl Nucl.~Phys.~{\bf B}}
\newcommand{\ptp}{\sl Prog.~Theor.~Phys.}
\newcommand{\ptps}{\sl Prog.~Theor.~Phys.~Suppl.}

\newcommand{\zpc}{\sl Z.~Phys.~{\bf C}}

\newcommand{\rmp}{\sl Rev.~Mod.~Phys.}
\newcommand{\mpla}{\sl Mod.~Phys.~Lett.~{\bf A}}

\newcommand{\ibid}{\it ibid.}

\begin{document}
\baselineskip 18pt

\begin{titlepage}
\preprintnumber{%
KEK-TH-567 \\
KEK Preprint 98-20 \\
TU-527 \\
RCNS-97-03 \\
Revised version \\
July 23, 1998
}
\vspace*{\titlesep}
\begin{center}
{\LARGE\bf
Flavor changing neutral current processes in $B$ and $K$ decays in the
supergravity model
}
\\
\vspace*{\titlesep}
{\large $^1$%
\footnote{%
Address from April 1, 1998: Theory Group, KEK, Tsukuba, Ibaraki, 305-0801 Japan.
}Toru Goto},
{\large $^2$Yasuhiro Okada,}
and
{\large $^2$Yasuhiro Shimizu}
\\
\vspace*{\authorsep}
{\it $^1~$Department of Physics, Tohoku University \\
  Sendai 980-8578 Japan}
\\
\vspace*{\authorsep}
{\it $^2$Theory Group, KEK, Tsukuba, Ibaraki, 305-0801 Japan }
\end{center}
\vspace*{\titlesep}
\begin{abstract}
Flavor changing neutral current processes such as \bsg, \bsll, \bsnn, 
\ek, \dmb, \kppnn\ and \klpnn\ are
calculated in the supersymmetric standard model based on supergravity.
We consider two assumptions for the soft supersymmetry breaking terms.
In the minimal case soft breaking terms for all scalar fields are taken
to be universal at the GUT scale whereas those terms are different for
the squark/slepton sector and the Higgs sector in the nonminimal case.
In the calculation we have taken into account the next-to-leading order
QCD correction to the \bsg\ branching ratio, the results from the LEP~II
superparticles search, and the condition of the radiative electroweak
symmetry breaking.
We show that \dmb\ and \ek\ can be enhanced up to 40\% compared
to the Standard Model values in the nonminimal case.
In the same parameter region the \bsnn, \kppnn\ and \klpnn\ branching
ratios are reduced up to 10\%. The corresponding deviation in the
minimal case is 20\% for \dmb\ and \ek\ and within 3\% for the
\bsnn, \kppnn\ and \klpnn.
For the \bsll\ process the significant deviation from the Standard Model
is realized only when the \bsg\ amplitude has an opposite sign to the
Standard Model prediction.
Significance on these results from possible future improvements of the
\bsg\ branching ratio measurement and top squark search is discussed.
\end{abstract}

\end{titlepage}

\section{Introduction}

Rare processes such as flavor changing neutral current (FCNC) processes
have been useful probes for the physics beyond the energy
scale directly accessible in collider experiments.
Among new physics beyond the standard model (SM), supersymmetry (SUSY)
is considered to be the most promising candidate.
Since FCNC is absent at the tree level in the minimal
supersymmetric standard model (MSSM) as in the SM, these rare processes 
can give useful constraints on the masses and mixings of the SUSY 
particles through loop diagrams.

Although squark masses are free parameters within the framework of the
MSSM, it is known that too large FCNC's are induced if we allow
arbitrary mass splittings and mixings among the
squarks with the same quantum numbers \cite{FCNC}.
This suggests that the SUSY breaking in the MSSM sector is induced 
from a generation-independent interaction.
A simple realization of the generation-independent SUSY breaking is the
minimal supergravity model. In this case the SUSY breaking in the hidden
sector is transmitted to the MSSM sector by the gravitational
interaction which does not distinguish the generation nor other gauge
quantum numbers. 
As a result, induced soft SUSY breaking masses are equal at the Planck
scale for all scalar fields in the MSSM sector.
FCNC processes have been studied extensively in the supergravity model as 
well as more general context of the SUSY models for the \kk\ and the
\bb\ mixings \cite{bbbar,BCKO,GM-BG,BBMR,GNO}, \bsg\
\cite{GM-BG,BBMR,bsg}, \bsll\ \cite{BBMR,GOST,bsll}, \bsnn\
\cite{BBMR,GOST} and \kpnn\ \cite{kpnn,kpnnparam}.
In Ref.~\cite{GNO} the \bb\ mixing and \ek\ (CP violating parameter in
the \kk\ mixing) were calculated in the minimal supergravity model under
the LEP constraints and it was shown that these quantities can be larger
than the SM values by 20\%.
Rare $b$ decay processes such as \bsg, \bsll, and \bsnn\ are considered
in Ref.~\cite{GOST} and it was pointed out that, taking account of the
LEP~1.5 constraints, there is a parameter region where
the \bsll\ branching ratio can be enhanced by 50\% compared to the SM 
value.  Also the \bsnn\ branching ratio is shown to be reduced at most 
by 10\% from the SM prediction.

In this way effects of SUSY particles and the charged Higgs boson 
vary from a few \% to several ten's \% depending on various FCNC 
processes. Since future experiments on $B$ and $K$ decays may reveal
new physics effects of this magnitude it is important to make
quantitative predictions using updated constraints on SUSY
parameter space.
Recent important theoretical improvement in this aspect is that the
complete next-to-leading order formula of the QCD correction to the
branching ratio of \bsg\ becomes available for the SM
\cite{bsgnloSManom}
and the two Higgs doublet models \cite{bsgnlo2H}.
As a result, the theoretical uncertainty in the calculation of
$\br(\bsg)$ has been reduced to $\lsim 10$\% level.

In this paper we study the SUSY contributions to FCNC processes under
the updated constraints.
We take account of the next-to-leading order QCD corrections for the
evaluation of $\br(\bsg)$ as well as the bounds on SUSY particle masses
from the recent LEP~II results \cite{LEPII} in order to obtain the
allowed region in the SUSY parameter space.
Then we evaluate various FCNC quantities such as \bsll, \bsnn, \bb\
mixing amplitude, \ek, \kppnn\ and \klpnn\ within the allowed parameter
region.
The numerical results depend on assumption of SUSY breaking terms at the
GUT scale.
In particular, in the minimal supergravity model soft SUSY breaking
terms for all scalar fields are assumed to be the same at the GUT scale.
If we would like to suppress too large SUSY contributions to 
the \kk\ mixing it is sufficient to require the degeneracy of 
the soft SUSY breaking masses only in the squark sector.
Because the strict universality for all scalar masses is not necessarily 
required in the context of the supergravity model, we study how the
allowed deviations of the FCNC quantities change when the universality
condition is relaxed.
For this purpose we take the soft SUSY breaking term for the Higgs
masses as a parameter independent of the universal squark/slepton mass.
This kind of assumption was considered in Ref.~\cite{MN} in a
different context.
We will see that the SUSY effects are considerably enhanced in a
parameter space which is excluded in the minimal case from the condition
of the proper electroweak symmetry breaking.
In the nonminimal case, the  branching ratios of \kpnn\ can be smaller 
than the SM values by 10\%, and at the same time, \ek\ and the 
\bb\ mixing become larger than the SM values by 40\% 
for $\tan\beta=2$.
The corresponding values in the minimal case are given by 3\% and 20\%,
respectively.
For \bsll, the result does not significantly differ from the minimal
case: there is a parameter space where the branching ratio becomes
larger by 50\% than the SM value for a large $\tan\beta$.
We analyze the correlation between the SUSY contributions to the FCNC
processes and the \bsg\ branching ratio.
It turns out that the maximal deviation occurs in the case that the
\bsg\ branching ratio is away from the central value of the SM
prediction.
We also show that the large deviation occurs in a parameter region where 
the top squark is relatively light.
Therefore the improvement in the \bsg\ branching ratio measurement and
the top squark mass bound will give great impacts on the SUSY search
through the various FCNC processes.

The rest of this paper is organized as follows.
In the next section, we introduce the supergravity model.
In Sec.~\ref{sec:fcnc} we describe the calculation of each FCNC
quantity.
In Sec.~\ref{sec:results}, our results of numerical analyses are
presented.
Sec.~\ref{sec:conclusions} is devoted for discussions and conclusions.

\section{Supergravity model}
\label{sec:model}

In this section we briefly outline calculations of the SUSY particles'
masses and the mixing parameters in the supergravity model for the
minimal and the nonminimal cases.
The actual procedure is the same as those discussed in
Ref.~\cite{GNA,GNO,GOST} except for the choice of the initial soft SUSY
breaking parameters for the nonminimal case.

The MSSM Lagrangian is specified by the superpotential and the soft SUSY 
breaking terms.
The superpotential is given by
\begin{eqnarray}
  W_{\rm MSSM}
  &=& f_D^{ij} Q_i D_j H_1
    + f_U^{ij} Q_i U_j H_2
    + f_L^{ij} E_i L_j H_1
    + \mu H_1 H_2 ~,
  \label{superpotential}
\end{eqnarray}
where the chiral superfields $Q$, $D$, $U$, $L$, $E$, $H_1$ and $H_2$
transform under $SU(3) \times SU(2) \times U(1)$ group as following
representations:
\begin{eqnarray}
  &&Q_i = (3           ,\,2,\, \frac{1}{6}) ~,~~
    U_i = (\overline{3},\,1,\,-\frac{2}{3}) ~,~~
    D_i = (\overline{3},\,1,\, \frac{1}{3}) ~,
\nonumber\\
  &&L_i = (1           ,\,2,\,-\frac{1}{2}) ~,~~
    E_i = (1           ,\,1,\, 1          ) ~,
\nonumber\\
  &&H_1 = (1           ,\,2,\,-\frac{1}{2}) ~,~~
    H_2 = (1           ,\,2,\, \frac{1}{2}) ~.
  \label{superfield}
\end{eqnarray}
The suffices $i,j=1,2,3$ are generation indices.
$SU(3)$ and $SU(2)$ indices are suppressed for simplicity.
A general form of the soft SUSY breaking terms is given by
\begin{eqnarray}
  -{\cal L}_{\rm soft}
  &=& (m_Q^2)^i_{~j} \wt{q}_i \wt{q}^{\dagger j}
    + (m_D^2)_i^{~j} \wt{d}^{\dagger i} \wt{d}_j
    + (m_U^2)_i^{~j} \wt{u}^{\dagger i} \wt{u}_j
  \nonumber\\
  &&+ (m_E^2)^i_{~j} \wt{e}_i \wt{e}^{\dagger j} 
    + (m_L^2)_i^{~j} \wt{l}^{\dagger i} \wt{l}_j
  \nonumber\\
  &&+ \Delta_1^2 h_1^\dagger h_1
    + \Delta_2^2 h_2^\dagger h_2
    - \left( B\mu h_1 h_2 + \hc \right)
  \nonumber\\
  &&+ \left(   A_D^{ij} \wt{q}_i \wt{d}_j h_1
             + A_U^{ij} \wt{q}_i \wt{u}_j h_2
             + A_L^{ij} \wt{e}_i \wt{l}_j h_1
             + \hc \right)
  \nonumber\\
  &&+ \left(   \frac{M_1}{2} \wt{B}\wt{B}
             + \frac{M_2}{2} \wt{W}\wt{W}
             + \frac{M_3}{2} \wt{G}\wt{G} + \hc \right) ~,
  \label{soft}
\end{eqnarray}
where $\wt{q}_i$, $\wt{u}_i$, $\wt{d}_i$,
$\wt{l}_i$, $\wt{e}_i$, $h_1$ and $h_2$  
are scalar components of chiral superfields
$Q_i$, $U_i$, $D_i$,
$L_i$, $E_i$, $H_1$ and $H_2$, respectively, and $\wt{B}$,
$\wt{W}$ and $\wt{G}$ are $U(1)$, $SU(2)$ and $SU(3)$ gauge
fermions, respectively.

In the framework of the supergravity model, the soft SUSY breaking
parameters are assumed to have a simple structure at the Planck scale.
In the present analysis, we take the following initial conditions at the
GUT scale $M_{\rm GUT} \sim 2 \times 10^{16}$ GeV.
We neglect the difference between the Planck and the GUT scales:
\begin{subeqnarray}
  (m_Q^2)^i_{~j} &=&
  (m_E^2)^i_{~j} ~=~ m_0^2\,\delta^i_{~j} ~,
\nonumber\\  
  (m_D^2)_i^{~j} &=&
  (m_U^2)_i^{~j} ~=~
  (m_L^2)_i^{~j} ~=~ m_0^2\,\delta_i^{~j} ~,
\\
  \Delta_1^2 &=& \Delta_2^2 ~=~ \Delta_0^2 ~,
\\
  A_D^{ij} &=& f_{DX}^{ij} A_X m_0 ~, ~~ 
  A_L^{ij} ~=~ f_{LX}^{ij} A_X m_0 ~, ~~ 
  A_U^{ij} ~=~ f_{UX}^{ij} A_X m_0 ~,
\\
  M_1 &=& M_2 ~=~ M_3 ~=~ M_{gX} ~.
\label{softGUT}
\end{subeqnarray}
In the minimal case $m_0$ and $\Delta_0$ are assumed to be equal,
whereas in the nonminimal case we take $m_0$ and $\Delta_0$ as
independent parameters.
We also assume that $A_X$, $M_{gX}$ and $\mu$ are all real parameters to
avoid a large electric dipole moment of the neutron \cite{edm}.
Therefore, no new CP violating complex phase (other than that in the
Cabibbo-Kobayashi-Maskawa (CKM) matrix) is introduced in the present
analysis. 

The soft SUSY breaking parameters at the electroweak scale are
calculated by solving the renormalization group equations (RGEs) of the
MSSM \cite{RGE} and we also impose the radiative electroweak symmetry
breaking condition \cite{REWSB}.
Taking the quark masses, the CKM matrix and
$\tan\beta =\vev{h_2^0}/\vev{h_1^0}$ as inputs,
we first solve one-loop RGEs for the gauge and Yukawa coupling constants
to calculate the values at the GUT scale.
Then we solve the RGEs for all MSSM parameters downward with initial
conditions \refeq{softGUT} for each set of the universal
soft SUSY breaking parameters $(m_0,\, \Delta_0,\, A_X,\, M_{gX})$.
We include all generation mixings in the RGEs for both Yukawa coupling
constants and the soft SUSY breaking parameters.
Next, we evaluate the Higgs potential at $m_Z$ scale including the
one-loop corrections induced by the Yukawa couplings constants of the
third generation \cite{Vloop}, and require that the minimum of the potential
gives correct vacuum expectation values of the neutral Higgs fields as
$\vev{h_1^0} = v\cos\beta$ and $\vev{h_2^0} = v\sin\beta$ where $v =
174$ GeV.
The requirement of the radiative electroweak symmetry breaking
fixes the magnitude of the SUSY Higgs mass parameter $\mu$ and the
soft SUSY breaking parameter $B$.
At this stage, all MSSM parameters at the electroweak scale are
determined as functions of the input parameters
$(\tan\beta,\, m_0,\, \Delta_0,\, A_X,\, M_{gX},\, \sign(\mu))$.
With use of the MSSM parameters at the electroweak scale,
we obtain the masses and the mixing parameters
(both angles and phases) of all the SUSY particles by diagonalizing the
mass matrices.
We impose the following phenomenological constraints on the
obtained particle spectra.
\renewcommand{\labelenumi}{\theenumi.} 
\renewcommand{\theenumi}{\arabic{enumi}} 
\begin{enumerate}
\item \bsg\ constraint from CLEO, \ie,
  $1.0 \times 10^{-4} < \br(\bsg) < 4.2 \times 10^{-4}$ \cite{CLEO}.
\item The chargino mass is larger than 91 GeV, and all other charged
  SUSY particle masses should be larger than 80 GeV \cite{LEPII}.
\item All sneutrino masses are larger than 41 GeV \cite{PDG}.
\item The gluino and squark mass bounds from TEVATRON experiments
  \cite{TEV}.
  The precise bounds on the gluino and squark masses depend on various
  SUSY parameters.
  Here we impose the constraint reported in Ref.~\cite{TEV} on the
  parameter space of the gluino mass and the averaged squark mass except
  for the top squark.
  Actually the gluino mass and the squark masses are more strictly
  constrained in this model from the chargino mass bound and the GUT
  relation of the gaugino masses, so that these masses are restricted to 
  be larger than about 200 GeV except for the lighter top squark.
  For the light top squark, the experimental bound is obtained at LEP
  and TEVATRON experiments \cite{lightstop} which was already taken into
  account in 2.
\item From the LEP neutralino search \cite{LEP},
  $\Gamma(Z\rightarrow\chi\chi) < 8.4$ MeV and 
  $\br(Z\rightarrow\chi\chi')$,
  $\br(Z\rightarrow\chi'\chi') < 2 \times 10^{-5}$ where $\chi$ is the
  lightest neutralino and $\chi'$ denotes other neutralinos.
\item The lightest SUSY particle is neutral.
\item The condition for not having a charge or color symmetry breaking
  minimum \cite{ccb}.
\end{enumerate}
In the next section we calculate the FCNC and/or CP violating quantities
such as the branching ratios for \bsll, \bsnn, \kppnn, \klpnn\ and 
the \bb\ mixing and \ek\ in the allowed parameter region.

\section{FCNC processes in $B$ and $K$ decays}
\label{sec:fcnc}

\subsection{\bsg, \bsll\ and \bsnn}
\label{sec:bsgbsllbsnn}

We basically follow Ref.~\cite{GOST} for the calculations of \bsg,
\bsll\ and \bsnn\ branching ratios, but we improve the calculation
taking into account the next-to-leading order QCD corrections.

The effective Hamiltonian for the $b \rightarrow s$ transition processes
is given as \cite{GSW,GOST,heff}
\begin{eqnarray}
{\cal H}^{\rm eff}_1 &=& \sum_{i=1}^{11} C_i(Q) {\cal O}_i(Q) + \hc ~,
\end{eqnarray}
where $Q$ is the renormalization point. 
The operators relevant to the present study are
\begin{subeqnarray}
{\cal O}_{ 7} &=& \frac{e}{(4\pi)^2} m_b
                  (\ol{s} \sigma^{\mu\nu} b_R)
                  F_{\mu\nu} ~,
\\
{\cal O}_{ 9} &=& \frac{e^2}{(4\pi)^2}
                  (\ol{s} \gamma^\mu b_L)
                  (\ol{l} \gamma_\mu l) ~,
\\
{\cal O}_{10} &=& \frac{e^2}{(4\pi)^2}
                  (\ol{s} \gamma^\mu b_L)
                  (\ol{l} \gamma_\mu \gamma_5 l) ~,
\end{subeqnarray}
for \bsg\ and \bsll, and
\begin{eqnarray}
{\cal O}_{11} &=& \frac{e^2}{(4\pi)^2}
                  (\ol{s} \gamma^\mu b_L)
                  (\ol{\nu} \gamma_\mu (1-\gamma_5) \nu) ~,
\end{eqnarray}
for \bsnn.
Other operators (the four-quark operators ${\cal O}_{1,2,\cdots,6}$ and
the chromomagnetic operator ${\cal O}_{8}$) contribute through the QCD
corrections.
We first calculate the Wilson coefficients $C_i$ at the electroweak
scale with use of the masses and the mixings of SUSY particles as well
as the SM ones.
Then we evaluate $C_i$ at $m_b$ scale including the QCD corrections
below the electroweak scale in order to obtain the branching ratios of
$b \rightarrow s$ decays.

As for the next-to-leading order QCD correction in the calculation of
$\br(\bsg)$, we follow
Ref.~\cite{AY,BKP,bsgnloSManom,bsgnloSMmatel,bsgnloSMbrems} for the SM
contribution and Ref.~\cite{bsgnlo2H} for the charged Higgs boson
contribution.
The QCD correction consists of the $O(\alpha_s)$ matching at the
electroweak scale \cite{AY,BKP,bsgnlo2H}, the next-to-leading order
anomalous dimension \cite{bsgnloSManom}, two-loop matrix elements
\cite{bsgnloSMmatel} and the Bremsstrahlung corrections
\cite{bsgnloSMbrems}.
In Ref.~\cite{BKP}, the SM value is given as
$\br(\bsg)_{\rm SM}^{\rm NLO} = ( 3.60 \pm 0.33 ) \times 10^{-4}$
compared to the leading order result
$\br(\bsg)_{\rm SM}^{\rm LO} = ( 2.8 \pm 0.8 ) \times 10^{-4}$.
$O(\alpha_s)$ matching conditions for the SUSY loop corrections have not 
been completed.
In Ref.~\cite{Anlauf}, these corrections are given for the case that the
ratio of the chargino mass and the top squark mass is large.
Since we are mainly interested in the case that both particles are
relatively light, we do not include these corrections.
Recently electroweak radiative corrections to $\br(\bsg)$ is considered
in Ref.~\cite{bsgEW}.
We will discuss these effects on the numerical results later although we 
have not included them explicitly in the calculation.
For the next-to-leading order QCD corrections to \bsll\ and \bsnn\ we
follow Ref.~\cite{bsllnlo,heff}.

The main SM contributions to the $b \rightarrow s$ decays come from the
loop diagrams involving the top quark and the relevant CKM matrix element
is $V_{ts}^* V_{tb}$, which is approximately written as $V_{ts}^* V_{tb} 
\approx -V_{cs}^* V_{cb}$ because of the unitarity and the smallness of
$V_{us}^* V_{ub}$.
Also the charm quark loop contribution has the CKM factor $V_{cs}^*
V_{cb}$.
Consequently, unlike \bb\ mixing, \ek\ and \kpnn\ the SM values of the
branching ratios for above processes are calculable without much
uncertainty since the relevant CKM factors are known in a good accuracy.

The SM predictions of the branching ratios for these processes are
$\br(\bsll)\simeq 0.8(0.6)\times 10^{-5}$ for $l=e\, (\mu)$ and 
$\br(\bsnn)\simeq 4.2 \times 10^{-5}$.
These processes have not yet observed experimentally and
only upper bounds are given by
$\br(\bsll)< 5.7(5.8)\times 10^{-5}$ for $l=e\, (\mu)$ \cite{CLEO2} and
$\br(\bsnn)<3.9 \times 10^{-4}$ \cite{GLN}.
The \bsll\ process is expected to be observed in the near future at the
$B$ factories and hadron machines.

\subsection{\kppnn\ and \klpnn}

The branching ratios of \kpnn\ processes are calculated by evaluating
the Wilson coefficient $C_{11}^d$ in the effective Hamiltonian
\begin{eqnarray}
{\cal H}^{\rm eff}_d &=& C_{11}^d {\cal O}_{11}^d + \hc ~,
\nonumber\\
{\cal O}_{11}^d &=& \frac{e^2}{(4\pi)^2}
                  (\ol{s} \gamma^\mu d_L)
                  (\ol{\nu} \gamma_\mu (1-\gamma_5) \nu) ~,
\end{eqnarray}
in a similar way as \bsnn.
The branching ratios normalized to that of the $K_{e3}$ decay are
written as
\begin{subeqnarray}
\label{eq:kpnn}
  \frac{ \br(\kppnn) }{ \br(\kppen) } &=&
  \left( \frac{\alpha}{4\pi} \right)^2
  \frac{ \sum_{\nu} \left| C_{11}^{d} \right|^2 }%
       { \left| V_{us} \right|^2 G_F^2 }
  r_{K^+} ~,
\\
  \frac{ \br(\klpnn) }{ \br(\kppen) } &=&
  \left( \frac{\alpha}{4\pi} \right)^2 
  \frac{ \sum_{\nu}\left| \im C_{11}^{d} \right|^2 }%
       { \left| V_{us} \right|^2 G_F^2 }
  \frac{ \tau_{K_L} }{ \tau_{K^+} }
  r_{K_L} ~,
\end{subeqnarray}
where $\tau_{K_L}(\tau_{K^+})$ denotes the lifetime for $K_L(K^+)$
and $r_{K^+}$ and $r_{K_L}$ are isospin breaking factors \cite{MP}.

The SM contributions to $C_{11}^{d}$ come from both the top and the
charm loops with CKM factors $V_{ts}^*V_{td}$ and $V_{cs}^*V_{cd}$,
respectively.
The dependencies on $V_{td}$ (or $\rho$ and $\eta$ in the Wolfenstein's
parametrization) are different in \kppnn\ and \klpnn\ since only the
$V_{ts}^*V_{td}$ term contributes to \klpnn\ while the sum of both terms
contributes to \kppnn.
The details of the calculation of \kpnn\ processes in the SM are
available in Ref.~\cite{heff}.
Following this reference, we have taken into account the
next-to-leading order QCD correction to the SM contribution.

The SM predictions for above branching ratios are given by
$\br(\kppnn)=(0.6\mbox{--}1.5)\times 10^{-10}$ and
$\br(\klpnn)=(1.1\mbox{--}5.0)\times 10^{-11}$
taking into account the ambiguity of unknown CKM parameters \cite{heff}.
Recently one candidate event of \kppnn\ is reported and the branching
ratio derived from this observation corresponds to
$4.2^{+9.7}_{-3.5}\times 10^{-10}$ \cite{E787}.
On the other hand for \klpnn\ only the upper bound is given by 
$\br(\klpnn)< 1.8\times 10^{-6}$ \cite{KTEV}.
Although the upper bound is still $10^5$ larger than the SM prediction,
dedicated searches for \klpnn\ are planned at KEK \cite{KEK}, BNL
\cite{BNL} and Fermilab \cite{KAMI}.
The \kpnn\ processes are theoretically very clean and 
the theoretical errors, such as QCD corrections, are expected to be
$\lsim 10$\% for \kppnn\ and  a few \% for \klpnn\ \cite{heff}.
Therefore \kpnn\ processes may give useful information on
the SUSY parameters if the branching ratios are measured at 10\% level.

\subsection{\bb\ mixing and \ek}

The \bb\ mixing matrix element $M_{12}(B)$ is calculated from the
effective Hamiltonian
\begin{eqnarray}
{\cal H}^{\rm eff}_2 &=& \frac{1}{128\pi^2} A(B)
  ( \ol{d} \gamma^\mu b_L )
  ( \ol{d} \gamma_\mu b_L ) + \hc ~,
\end{eqnarray}
with
\begin{eqnarray}
  M_{12}(B) &=& \frac{1}{2m_B}
  \langle B^0 | {\cal H}^{\rm eff}_2 | \ol{B}^0 \rangle
\nonumber\\ &=&
  \frac{\hat{B}_B \eta_B f_B^2 m_B}{384\pi^2} A(B) ~,
\end{eqnarray}
where $m_B$, $f_B$, $\hat{B}_B$ and $\eta_B$ are the $B$-meson mass, decay
constant, bag parameter and QCD correction factor, respectively.
The \kk\ mixing matrix element $M_{12}(K)$ is obtained in the same way 
by replacing the external bottom quark with the strange quark and the
\ek\ is proportional to $\im M_{12}(K)$.
We calculate the coefficient $A(B)$ and $A(K)$ as described in
Ref.~\cite{GNO} with the inclusion of the next-to-leading order QCD
corrections given in Ref.~\cite{HN}.
The experimental values for the \bb\ and \kk\ mixings are given as
$\Delta m_B=2\left|M_{12}(B)\right|=(0.474\pm 0.031)\ \mbox{ps}^{-1}$
\cite{PDG,bbbarex} and $|\ek|=(2.280\pm 0.013)\times 10^{-3}$
\cite{PDG}. 
At present these observables do not constrain the SUSY parameters 
very strongly because the CKM parameters relevant to these quantities
are not well-determined and considerable hadronic uncertainties still
exist in $\hat{B}_K$, $\hat{B}_B$ and $f_B$.

\section{Numerical Results}
\label{sec:results}

In this section we show our numerical results.
We scan the soft SUSY breaking parameter space in the range of
$m_{0} \leq 600$ GeV, $\Delta_{0} \leq 600$ GeV, $M_{gX} \leq 600$ GeV
and $|A_X| \leq 5$ for each fixed value of $\tan\beta$.
For the CKM matrix, we use the `standard' phase convention of the
Particle Data Group \cite{PDG}, taking $V_{us} = 0.2205$,
$V_{cb}=0.041$, $|V_{ub}/V_{cb}| = 0.08$ and $\delta_{13} = 90^{\circ}$
as input parameters.
We also change the value of $\delta_{13}$ and comment on the results if
necessary.
We fix the pole masses of the top, bottom and charm quarks as 175 GeV,
4.8 GeV and 1.4 GeV, respectively.
We also take $\alpha_s(m_Z) = 0.118$.

Let us first discuss general features of the mass spectrum and the
generation mixings of squarks determined by RGEs.
\begin{enumerate}
\item The first and second generation squarks with the same gauge
quantum numbers remain highly degenerate in masses but the third
generation squarks, especially the top squark can be
significantly lighter due to the renormalization effect of the top
Yukawa coupling constant.
\item  The squark flavor mixing matrix which
diagonalize the squark mass matrix is approximately the same as
corresponding CKM matrix apart from the left-right mixing of the top
squarks.
\end{enumerate}
As a result, SUSY contributions to the $b \rightarrow s$
$(s \rightarrow d)$ transition amplitudes and $M_{12}(B)$ ($M_{12}(K)$) 
are proportional to $V_{tb}V_{ts}^\ast$ ($V_{ts}V_{td}^\ast$) and
$(V_{tb}V_{td}^\ast)^2$ ($(V_{ts}V_{td}^\ast)^2$), respectively. 
Therefore CP violating phase of $M_{12}(B(K))$ is equal to that in the 
SM. These features are the same as those in the minimal case
\cite{GNA,GNO,GOST}.

The quantitative difference between the minimal and the nonminimal
choices of the soft SUSY breaking parameters appears in the mass
spectrum.
In \reffig{fig:stp-cno} we show the allowed region in the space of the
lighter chargino and the lighter top squark masses for a different
assumption on $m_0$ and $\Delta_{0}$ for $\tan\beta=2$ and 30. 
We present the allowed region for the full parameter space, and the
minimal case ($m_0=\Delta_0$).
Contrary to the minimal case we see that a relatively light top squark
and chargino with masses $m_{\wt{t}_1} \sim 100$ GeV and 
$m_{\wt{\chi}^\pm_1} \sim 100$ GeV are simultaneously realized
especially for $\tan\beta = 2$.
This difference of the allowed mass spectrum leads to a quantitative
change in the prediction of the FCNC observables for the minimal and the 
nonminimal cases.

\subsection{\bsg, \bsll\ and \bsnn}

As discussed above the SUSY contribution to the $b \rightarrow s$
transition amplitudes is proportional to the $V_{tb}V_{ts}^\ast$ element
just as the SM and the charged Higgs boson contributions.
As discussed in the subsection \ref{sec:bsgbsllbsnn},
the $V_{tb}V_{ts}^\ast$ element is well-constrained from the unitarity of
the CKM matrix so that there is little ambiguity associated with this
input parameter.
The Wilson coefficients $C_{7}$, $C_{9}$ and $C_{10}$ are 
relevant to the \bsg\ and \bsll\ decays.
The values of $C_{7}$, $C_{9}$ and $C_{10}$ in the supergravity model
are shown in \reffig{fig:c9c10-c7}.
Each coefficient is evaluated at the bottom mass scale and is normalized
by the corresponding SM value.
The SUSY contribution to $C_{7}$ can be as large as or even lager than
the SM contribution especially for a large $\tan\beta$.
We can see that the sign of $C_{7}$ can be opposite to that of the SM
prediction. On the other hand the SUSY contributions to $C_{9}$ and
$C_{10}$ are relatively small and interfere with SM ones constructively
in $C_{9}$ and destructively in $C_{10}$.
These features are the same as those in the minimal case discussed in
Ref.~\cite{GOST}.

In \reffig{fig:bsg-cH} we show the branching ratio of \bsg\ as a
function of the charged Higgs boson mass for $\tan\beta = 2$ (minimal
and nonminimal cases) and $\tan\beta = 30$ (nonminimal case).
For $\tan\beta = 30$, the plot looks the same even if the parameter
space is restricted to the minimal case.
Here we fix the renormalization point $\mu_b$ as $\mu_b = m_b$.
In the calculation of $\br(\bsg)$ we use the electromagnetic coupling
constant $\alpha_{\rm EM}$ at $m_b$ scale which is given by
$\alpha_{\rm EM}^{-1}(m_b) \simeq 132.3$.
Considering that the next-to-leading order formulas still contain
theoretical ambiguities due to the $\mu_b$ dependence and the choice of
the various input parameters, we should allow theoretical uncertainty at
10\% level for each point.
It is interesting to notice that for the minimal case with
$\tan\beta = 2$ there are two branches for $\br(\bsg)$.
In one branch the branching ratio is close to the two Higgs doublet
model (type II) prediction, therefore the contributions from SUSY
particles are small.
In the other branch it is consistent with the SM value, so that the
charged Higgs boson contribution is canceled by the SUSY contributions.

In \reffig{fig:bsll-bsg} we show the correlation between the branching 
ratios of \bsg\ and \bsmm. 
In this figure in order to avoid the $J/\psi$ resonance we use the
branching ratio for \bsmm\ integrated in the
region $2m_\mu<\sqrt{s}<m_{J/\psi}-100$ MeV where $\sqrt{s}$ is the
invariant mass of $\mu^+\mu^-$ pair.
As discussed in Ref.~\cite{GOST}, the branching ratio in this region
depends on the phase of the $b$--$s$--$J/\psi$ coupling $\kappa$ through 
the interference effect.
Although the branching ratio can change by $\pm 15$\%, this ambiguity
will be reduced if we can measure the lepton invariant mass spectrum
near the $J/\psi$ resonance region.
As an example we take $\kappa$ as $+1$ here.
We can see that a strong correlation between the two 
branching ratios since only $C_{7}$ receives the large SUSY
contribution.
In the present supergravity model therefore a large deviation of
$\br(\bsll)$ from the SM prediction is expected only when the
sign of $C_{7}$ is opposite to that in the SM, which is realized for a
large $\tan\beta$. 
This situation is similar to the minimal case \cite{GOST}.

The amplitude of \bsnn\ is determined by the Wilson coefficient $C_{11}$.
Apart from the CKM matrix element the SUSY contribution to $C_{11}$ is
the same as the SUSY contribution to $C_{11}^d$.
The branching ratio for \bsnn\ normalized by the SM prediction
($\br(\bsnn)/\br(\bsnn)_{\rm SM}$) is practically the same as 
a similar ratio for \klpnn\
($\br(\klpnn)/\br(\klpnn)_{\rm SM}$),
which is discussed in the next subsection.

\subsection{\kppnn\ and \klpnn}

As shown in \refeq{eq:kpnn} the branching ratios for \kppnn\ and \klpnn\
are proportional to $\left| C_{11}^{d} \right|^2$ and 
$\left| \im C_{11}^{d} \right|^2$, respectively.
In the SM $C_{11}^d$ is divided into two parts according to the
relevant CKM matrix elements as follows:
\begin{eqnarray}
  \label{c11d}
C_{11}^d = V_{td}V_{ts}^\ast C_{11}^d(\mbox{top}) 
+ V_{cd}V_{cs}^\ast C_{11}^d(\mbox{charm}).
\end{eqnarray}
As discussed before the SUSY contribution is proportional to 
$V_{td}V_{ts}^*$ therefore we can write 
\begin{eqnarray}
C_{11}^d \simeq V_{td}V_{ts}^\ast (C_{11}^d(\mbox{top})+C_{11}^d(\mbox{SUSY}))
+ V_{cd}V_{cs}^\ast C_{11}^d(\mbox{charm}),
\label{c11d2}
\end{eqnarray}
where $C_{11}^d(\mbox{SUSY})$ is the SUSY contribution including the
charged Higgs boson contribution.
This kind of parametrization for \kpnn\ is considered in Ref.~\cite{kpnnparam}.

In \reffig{fig:kpnn-cno} we show the branching ratio for \klpnn\
normalized by the SM prediction as a function of the lighter chargino
mass and the lighter top squark mass for $\tan\beta=2$.
Also the correlation with the $\br(\bsg)$ is shown.
In \reffig{fig:kpnn-cno}(a) and \reffig{fig:kpnn-cno}(b), we use the
CLEO bound on $\br(\bsg)$ as a constraint on the SUSY parameter space.
In order to take into account the theoretical ambiguity in a simple way, 
we allow 10\% uncertainty in the branching ratio and use
$(1.0 \times 10^{-4}) \times 0.9$ and $(4.2 \times 10^{-4}) \times 1.1$
as lower and upper bounds, respectively.
Note that the ratio $\br(\klpnn)/\br(\klpnn)_{\rm SM}$ does not depend
on the CKM parameters because only the first term in \refeq{c11d} 
contributes to this process.
We see that the branching ratio for \klpnn\ becomes smaller than
the SM prediction by 10\%.
In the minimal case the maximal deviation is within 3\%.
We investigated in which parameter region the maximal deviation is
realized.
We found that the large deviation occurs in the $m_0\simeq 150$
GeV and $\Delta_0\simeq 400$ GeV region which corresponds to the
parameter region with $m_{\chi^\pm_1},\,m_{{\wt t}_1}\simeq 100$ GeV
shown in \reffig{fig:stp-cno}.
From \reffig{fig:kpnn-cno}(c) we can see that a sizable reduction of
$\br(\klpnn)$ occurs when $\br(\bsg)$ becomes larger than the SM value.
We also calculate $\br(\klpnn)$ for different $\tan\beta$ and found that
the deviation becomes smaller for a large $\tan\beta$.
For example the maximal deviation is about 5\% for $\tan\beta=30$.
As we can see in \refeq{c11d2}, the branching ratio of \kppnn\ and
\klpnn\ have a strong correlation.
We show the correlation for three different values of $\delta_{13}$ in
\reffig{fig:klpnn-kppnn}.
In this figure we fix $m_0 = 150$ GeV, but the correlation does not
depend on the value of $m_0$.
The deviation from the SM value for $\br(\kppnn)$ is about 20\% smaller 
than that for $\br(\klpnn)$.

\subsection{\bb\ mixing and \ek}

Just as in the \klpnn\ and \kppnn\ case, the \bb\ mass splitting \dmb\
and \ek\ normalized to SM values are linearly correlated with each other
as noted in \cite{BCKO,GNO}.
We show the correlation for $\delta_{13} = 30^{\circ}$, $90^{\circ}$ and 
$150^{\circ}$ in \reffig{fig:ek-dmb}.
We see that the deviation from SM in \ek\ is about 80\% of that in
\dmb.
In the following, we only show the results for \dmb, but the
corresponding results on \ek\ can be easily obtained from
\reffig{fig:ek-dmb}.
In \reffig{fig:dmb-cno} we show \dmb\ normalized by the SM value as a
function of the lighter chargino mass, the lighter top squark mass and
$\br(\bsg)$ for $\tan\beta=2$.
The deviation can be as large as 40\% in the nonminimal case whereas
20\% in the minimal case.
From \reffig{fig:dmb-cno}(b) we can see that the deviation larger than
20\% is realized only in the nonminimal case when the top squark mass
is smaller than 200 GeV.
In this region $\br(\bsg)$ also deviates from the SM value
significantly as shown in \reffig{fig:dmb-cno}(c).
This result indicates  the importance of the further improvement of the
$\br(\bsg)$ measurement and the top squark search.
If the lower bound for the top squark mass is raised to 200 GeV, the
maximal deviation of \dmb\ is reduced to 25\%.
On the other hand, if the \bsg\ branching ratio turns out to be close to 
the present upper or lower bound, \dmb\ and \ek\ might be significantly
enhanced.
We should notice that because the theoretical uncertainty is already
reduced to 10\% level the experimental determination of $\br(\bsg)$ at
that level will put a strong constraint on the SUSY parameter space.
We also calculated \dmb\ for $\tan\beta = 30$ and found that the
deviation from the SM value is less than 10\%.
In \reffig{fig:kpnn-dmb} we show the correlation between $\br(\klpnn)$
and \dmb.
For $\tan\beta=2$ we see a strong correlation between these two
quantities: $\br(\klpnn)$ is reduced by 10\% when \dmb\ is enhanced by
40\%.
We can also see the correlation for $\tan\beta=30$.
In this case \dmb\ can be enhanced by 10\% in the region where
$\br(\klpnn)$ is reduced by 5\%.

\section{Conclusions and discussions}
\label{sec:conclusions}

In this paper we have studied the FCNC processes of $B$ and $K$ mesons
in the minimal supergravity model and in the supergravity model with an
extended parameter space of the soft SUSY breaking parameters.
We take into account the recent mass bounds for SUSY particles at LEP~II 
and the next-to-leading order QCD corrections to various processes
including \bsg.

We find that the branching ratio for \bsll\ can be enhanced by about
50\% compared to the SM value for a large $\tan\beta$ when the sign of
$C_7$ becomes opposite to that of SM. 
For $\tan\beta = 2$, the \bsnn, \kppnn\ and \klpnn\ processes have
similar SUSY contributions and it turns out that these branching ratios
are reduced at most 10\% in the nonminimal case whereas less than 3\% in
the minimal case.
The \bb\ mixing and \ek\ are enhanced up to 40\% from the SUSY
contributions in the nonminimal case whereas 20\% in the minimal case.
We investigate the correlation among \dmb, \ek\ and $\br(\kpnn)$, and
found that the large deviation occurs when the chargino is lighter than
150 GeV and the top squark is lighter than 200 GeV.
In the same parameter region $\br(\bsg)$ is close to the upper or lower
bound of the presently allowed region.
For a large $\tan\beta$, the deviations of \dmb, \ek\ and $\br(\kpnn)$
are smaller.
In the minimal case these deviations are somewhat smaller than 
the previous calculation \cite{GNO,GOST} especially for \bsnn\ .
This is because the mass bounds for chargino \etc\ have been improved by 
the LEP~II experiments.
We note that the maximal deviation depends on the light top squark mass
bound.
Therefore  the light top squark search in TEVATRON experiments can reduce
a possible parameter space where a large deviation from the SM value in
FCNC processes is realized. 

In this paper we extend the minimal supergravity model by introducing an 
additional parameter for the soft SUSY breaking term in the Higgs
sector.
This is not the unique way to extend the soft SUSY breaking terms.
In order to avoid too large FCNCs, we only require that the
squarks/sleptons in the same quantum numbers should have the common mass
term at the Planck scale.
Since the main difference is the change of the SUSY mass spectrum, a
deviation with a similar magnitude is expected to be realized in a
more general case as long as a light top squark and  light chargino
mass region is allowed.

In Ref.~\cite{bsgEW} electroweak radiative corrections to $\br(\bsg)$ is 
computed.
They found that the fermion and the photonic loop effects reduce the
branching ratio by $9\pm2$\%.
It is argued that the dominant contribution is due to the electric
charge renormalization, and as a result the electromagnetic coupling
constant should be evaluated at $q^2=0$, \ie,
$\alpha^{-1}_{\rm EM}(0)=137.036$.
Since we use $\alpha_{\rm EM}(m_b)$, this correction reduces $\br(\bsg)$ 
by 3\%.
  
Let us finally discuss the implications of these results when various
information is obtained in future $B$ and $K$ decay experiments.
Firstly, since no new phase appears in $M_{12}(B)$, the CP asymmetry
measured in the $B^0(\ol{B}^0) \rightarrow J/\psi\,K_S$
decay is directly related to the angle
$\phi_1=\arg \left(-\frac{V_{td}^*V_{tb}}{V_{cd}^*V_{cb}}\right)$ of the
unitarity triangle.
CP asymmetries in other $B$ decay modes and the ratio of the \dmb's for
$B_s$ and $B_d$ also provide information on the CKM matrix elements as
in the SM. 
On the other hand, ``$|V_{td}|$'' obtained from \dmb\ and \ek\ may be
different from that obtained above if we assume the SM analysis.
In the same way ``$|V_{td}|$'' from the branching ratios of \klpnn\ and
\kppnn\ may be different.
As shown in \reffig{fig:kpnn-dmb}, the SUSY contributions are
constructive to the SM contribution in \dmb\ (\ek) and destructive in
$\br(\kpnn)$ so that the deviations of ``$|V_{td}|$'' from the true
value become opposite.
Therefore combining CP asymmetry in $B$ decay, \dmbs\ and various FCNC
observables in $B$ and $K$ decays, we may obtain a hint on the existence 
of SUSY particles.

\section*{Acknowledgments}

The work of Y.~O. was supported in part by the
Grant-in-Aid for Scientific Research from the Ministry of Education,
Science and Culture of Japan.
The work of T.~G. was supported in part by the Soryushi Shogakukai.

\newpage



\clearpage
\section*{Figure Captions}

\newcounter{FIG}
\begin{list}{{\bf FIG. \arabic{FIG}}}{\usecounter{FIG}}
\item
  Allowed regions in the space of the lighter chargino mass
  $m_{\wt{\chi}^\pm_1}$ and the lighter top squark mass $m_{\wt{t}_1}$ for
  (a) $\tan\beta = 2$ and
  (b) $\tan\beta = 30$.
  The dots represent the allowed region for the full parameter space and
  the squares show the allowed region for the minimal case
  ($m_0=\Delta_0$).
\label{fig:stp-cno}
\item
$C_{7}$, $C_{9}$ and $C_{10}$ normalized to the SM values for
  (a) the full parameter space with $\tan\beta = 2$;
  (b) the minimal case with $\tan\beta = 2$;
  (c) the full parameter space with $\tan\beta = 30$; and
  (d) the minimal case with $\tan\beta = 30$.
\label{fig:c9c10-c7}
\item
$\br(\bsg)$ in the supergravity model as a function of the
  charged Higgs mass for (a) $\tan\beta = 2$ and (b) $\tan\beta = 30$.
  Each solid line shows the branching ratio in the two Higgs doublet
  model (type II).
  Each dashed line shows the branching ratio in the SM.
  Dotted lines denote the upper and lower bounds on the branching ratio
  given by CLEO.
  For $\tan\beta = 2$ the values in the minimal case is also plotted
  with circles.
\label{fig:bsg-cH}
\item
Branching ratios of \bsg\ and \bsmm\ for
  (a) $\tan\beta = 2$; and
  (b) $\tan\beta = 30$.
  Here, $\br(\bsmm)$ is obtained by integrating in the range
  $2m_\mu<\sqrt{s}<m_{J/\psi}-100$ MeV where $\sqrt{s}$ is the invariant
  mass of $\mu^+\mu^-$ pair.  
  The dots show the values in the full parameter space, the squares
  show those in the minimal case and the circle represents the SM value.
  The vertical dotted lines show  the upper and lower bounds on
  $\br(\bsg)$ given by CLEO.
\label{fig:bsll-bsg}
\item
The branching ratio for \klpnn\ normalized to the SM value for
  $\tan\beta=2$
  (a) as a function of the lighter chargino mass;
  (b) as a function of the lighter top squark mass; and
  (c) as a function of $\br(\bsg)$.
  Each dot represents the value in the full parameter space and
  each square shows the value for the minimal case.
  The vertical dotted lines in (c) show  the upper and lower bounds on
  $\br(\bsg)$ given by CLEO.
  In (a) and (b) the CLEO bound is imposed (see text).
\label{fig:kpnn-cno}
\item
Correlation between
  $\br(\kppnn)\slash\br(\kppnn)_{\rm SM}$ and
  $\br(\klpnn)\slash\br(\klpnn)_{\rm SM}$ for $\tan\beta=2$.
  Here, $m_0$ is fixed to 150 GeV and $\delta_{13}$ is taken as
  $30^{\circ}$, $90^{\circ}$ and $150^{\circ}$.
\label{fig:klpnn-kppnn}
\item
Correlation between
  $\ek/(\ek)_{\rm SM}$ and $\dmb/(\dmb)_{\rm SM}$ for $\tan\beta=2$.
  Here, $m_0$ is fixed to 150 GeV and $\delta_{13}$ is taken as
  $30^{\circ}$, $90^{\circ}$ and $150^{\circ}$.
\label{fig:ek-dmb}
\item
\dmb\ normalized by the SM value for $\tan\beta=2$
  (a) as a function of the lighter chargino mass;
  (b) as a function of the lighter top squark mass; and
  (c) as a function of $\br(\bsg)$.
  Each dot represents the value in the full parameter space and
  each square shows the value for the minimal case.
  The vertical dotted lines in (c) show  the upper and lower bounds on
  $\br(\bsg)$ given by CLEO.
  In (a) and (b) the CLEO bound is imposed.
\label{fig:dmb-cno}
\item
Correlation between
  $\br(\klpnn)\slash\br(\klpnn)_{\rm SM}$ and
  $\dmb/(\dmb)_{\rm SM}$ for
  (a) $\tan\beta = 2$; and
  (b) $\tan\beta = 30$.
\label{fig:kpnn-dmb}
\end{list}


\clearpage
\section*{Figures}
\pagestyle{empty}

\def\EPSDIR{}
\def\EPSSCALE{1.0}

~
\vfill
\begin{center}
\makebox[0cm]{
\def\epsfsize#1#2{\EPSSCALE#1}
\epsfbox{\EPSDIR 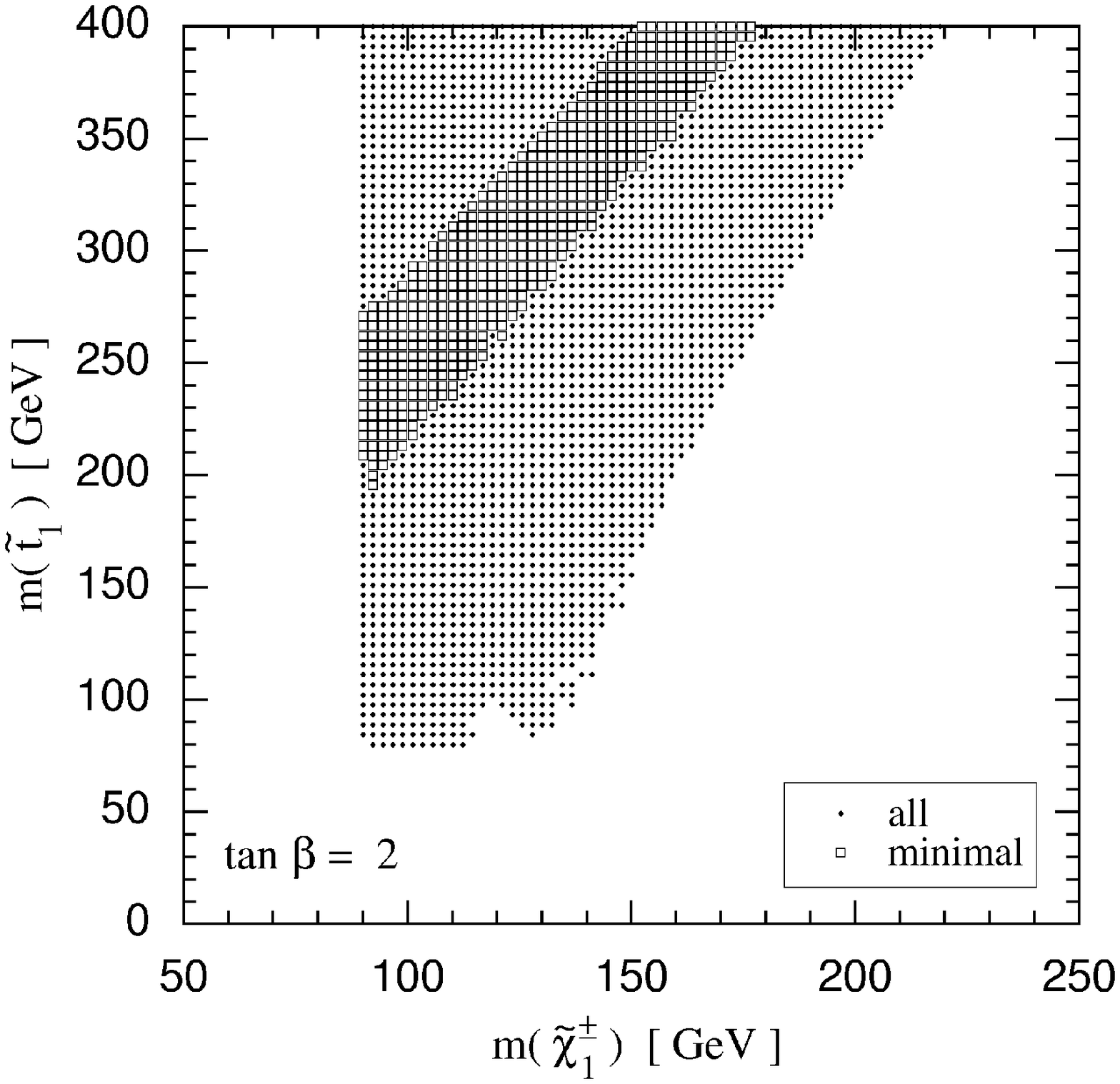}
}
\vfill
{\Large\bf Fig.~\ref{fig:stp-cno}(a)}
\end{center}
\clearpage

~
\vfill
\begin{center}
\makebox[0cm]{
\def\epsfsize#1#2{\EPSSCALE#1}
\epsfbox{\EPSDIR 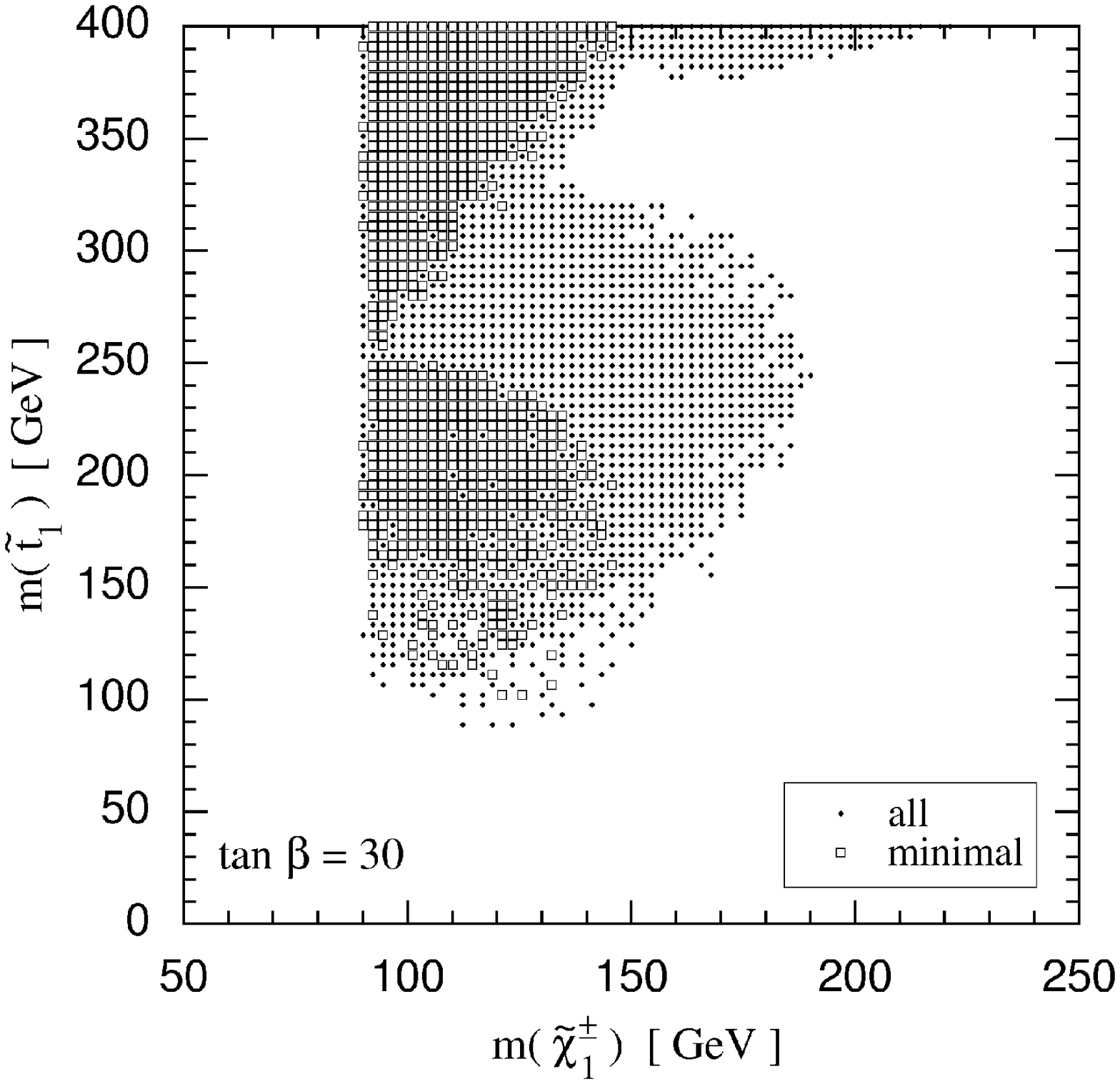}
}
\vfill
{\Large\bf Fig.~\ref{fig:stp-cno}(b)}
\end{center}
\clearpage

~
\vfill
\begin{center}
\makebox[0cm]{
\def\epsfsize#1#2{\EPSSCALE#1}
\epsfbox{\EPSDIR 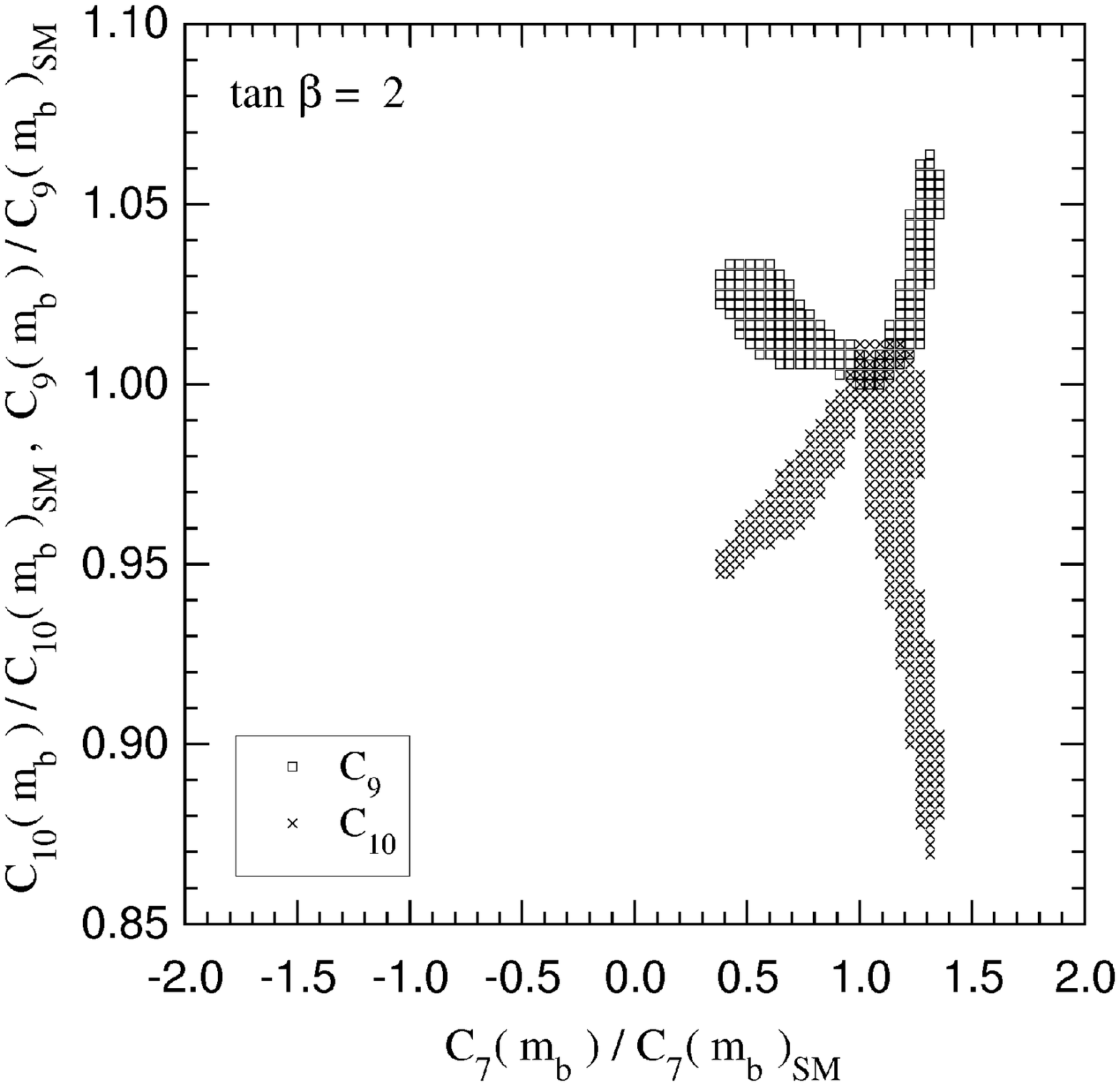}
}
\vfill
{\Large\bf Fig.~\ref{fig:c9c10-c7}(a)}
\end{center}
\clearpage

~
\vfill
\begin{center}
\makebox[0cm]{
\def\epsfsize#1#2{\EPSSCALE#1}
\epsfbox{\EPSDIR 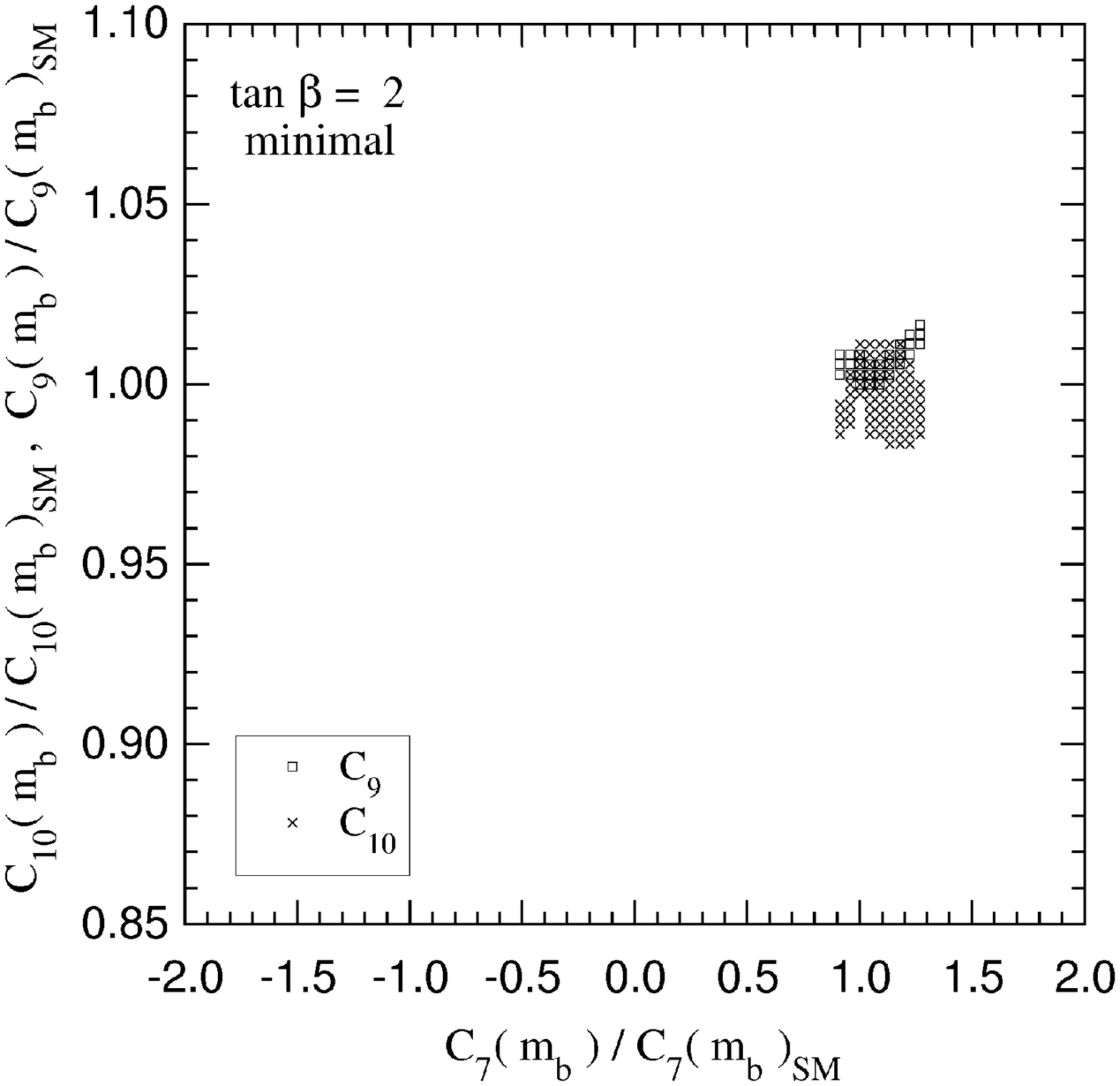}
}
\vfill
{\Large\bf Fig.~\ref{fig:c9c10-c7}(b)}
\end{center}
\clearpage

~
\vfill
\begin{center}
\makebox[0cm]{
\def\epsfsize#1#2{\EPSSCALE#1}
\epsfbox{\EPSDIR 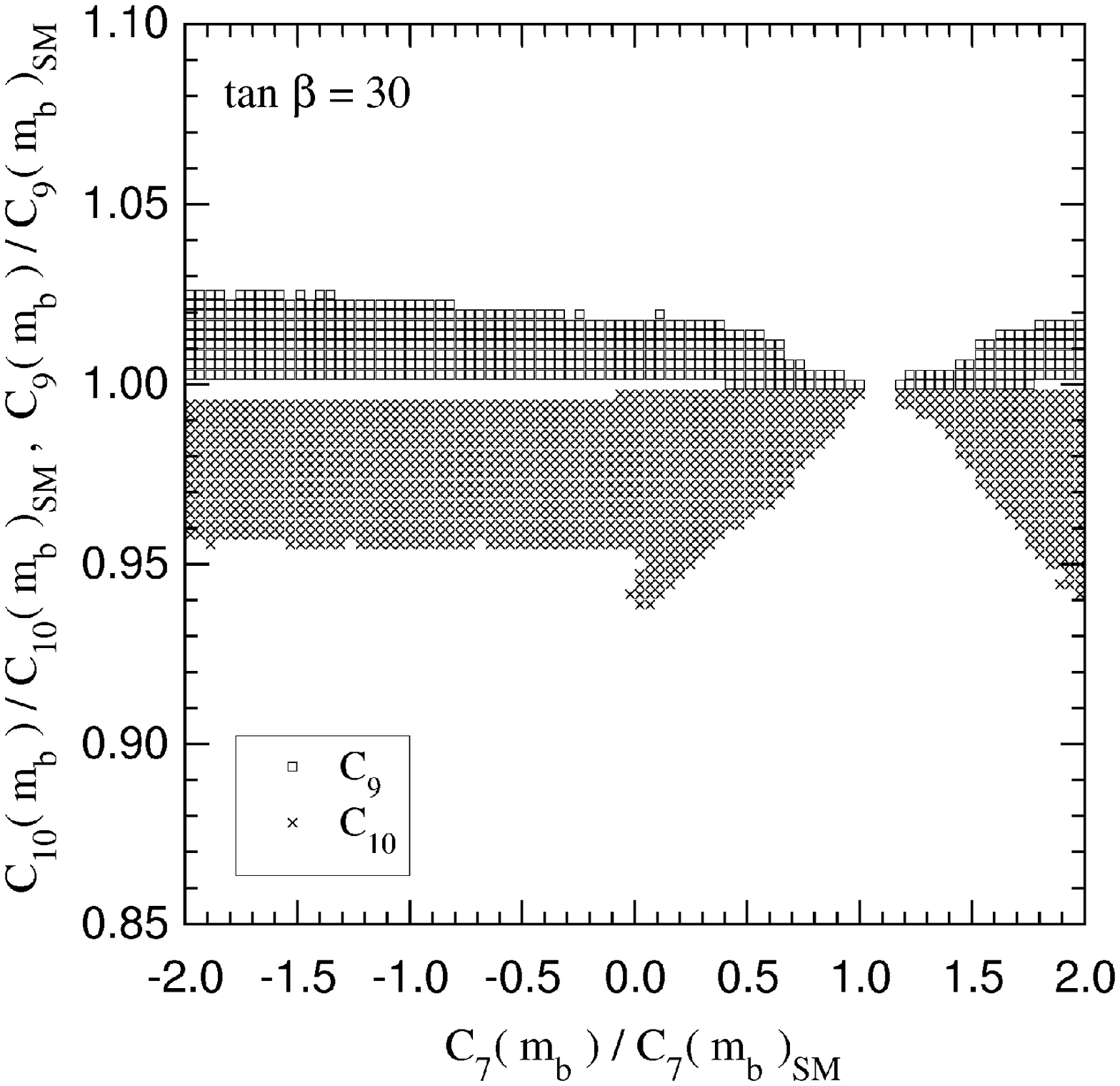}
}
\vfill
{\Large\bf Fig.~\ref{fig:c9c10-c7}(c)}
\end{center}
\clearpage

~
\vfill
\begin{center}
\makebox[0cm]{
\def\epsfsize#1#2{\EPSSCALE#1}
\epsfbox{\EPSDIR 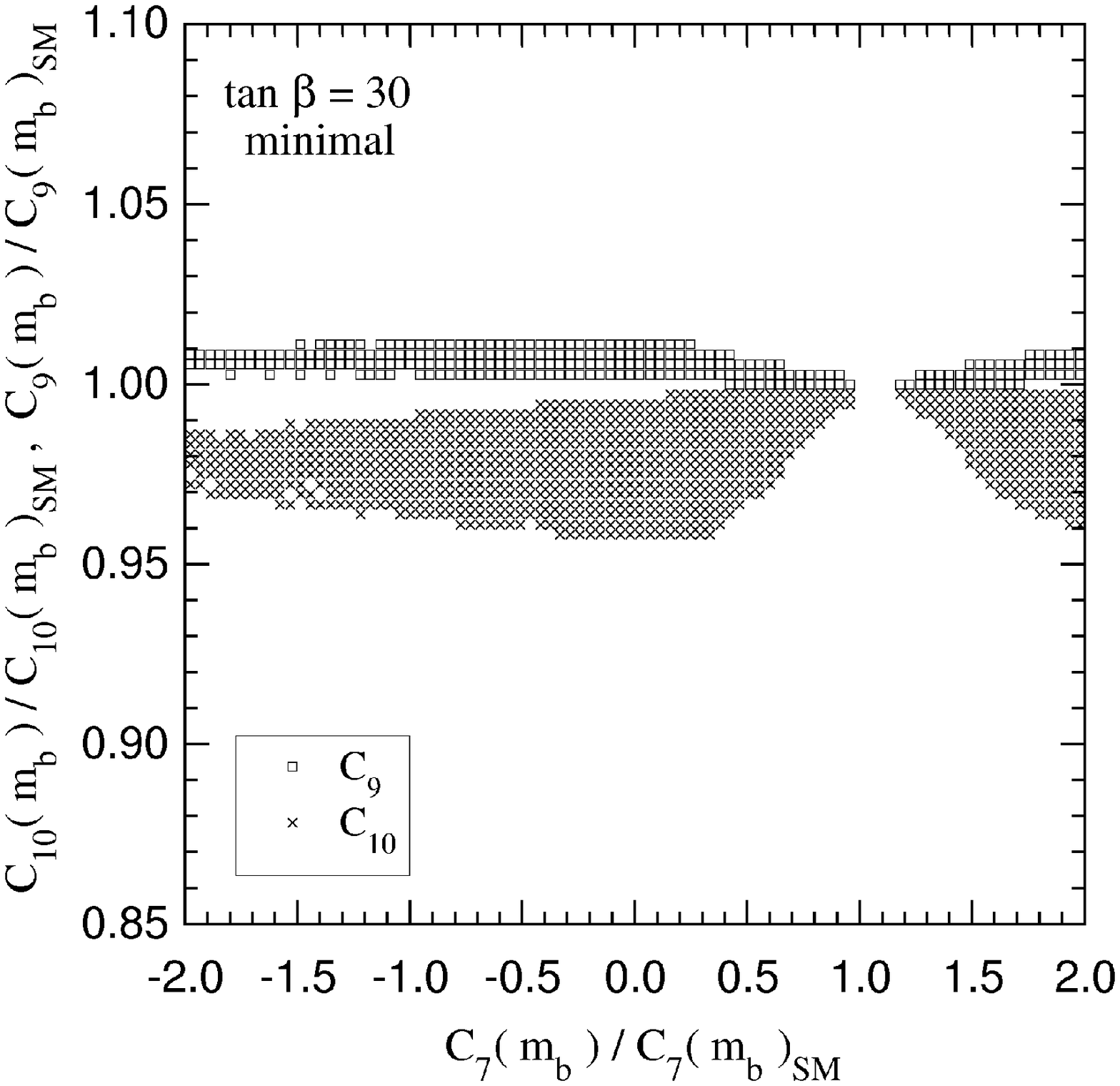}
}
\vfill
{\Large\bf Fig.~\ref{fig:c9c10-c7}(d)}
\end{center}
\clearpage

~
\vfill
\begin{center}
\makebox[0cm]{
\def\epsfsize#1#2{\EPSSCALE#1}
\epsfbox{\EPSDIR 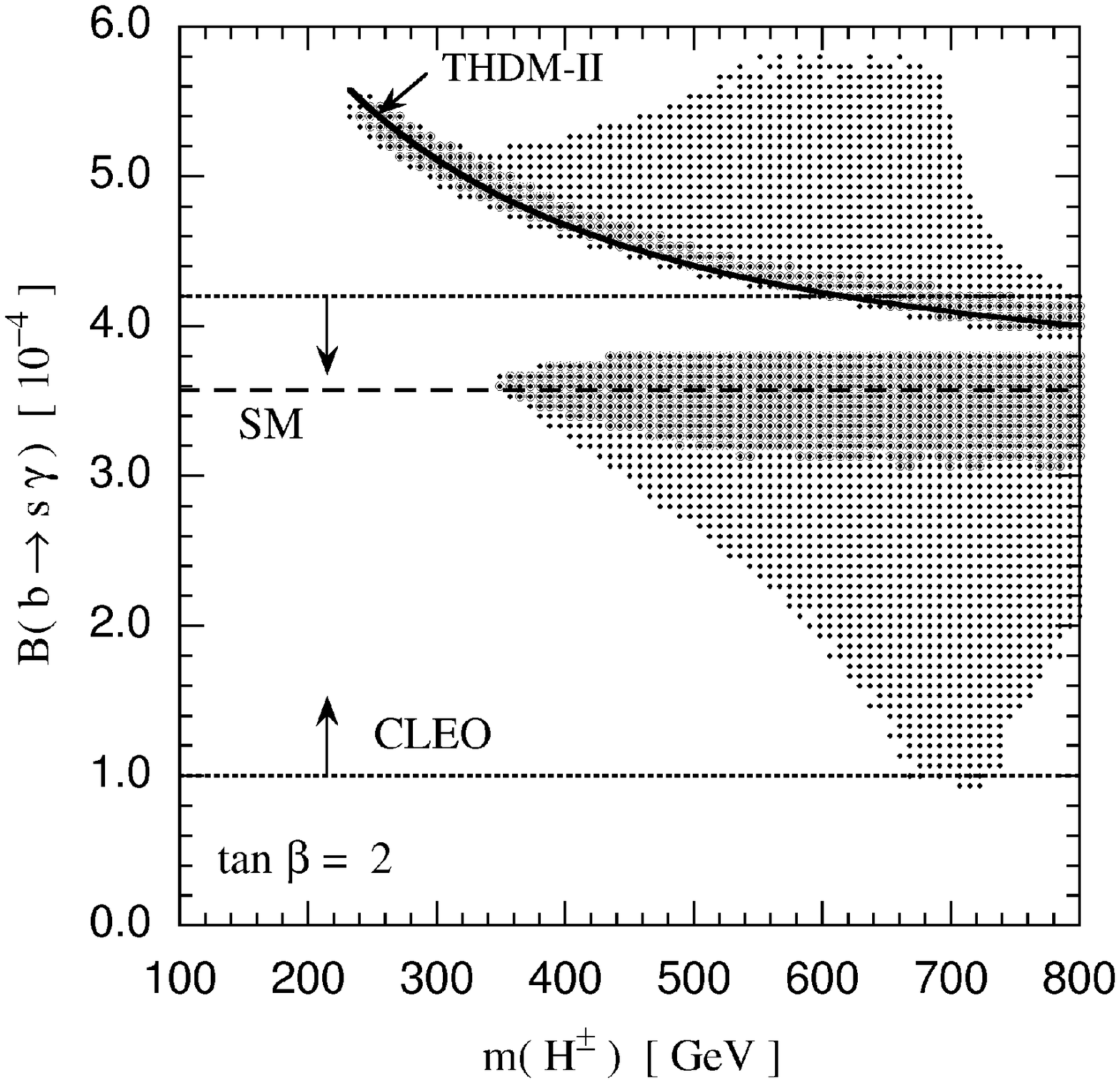}
}
\vfill
{\Large\bf Fig.~\ref{fig:bsg-cH}(a)}
\end{center}
\clearpage

~
\vfill
\begin{center}
\makebox[0cm]{
\def\epsfsize#1#2{\EPSSCALE#1}
\epsfbox{\EPSDIR 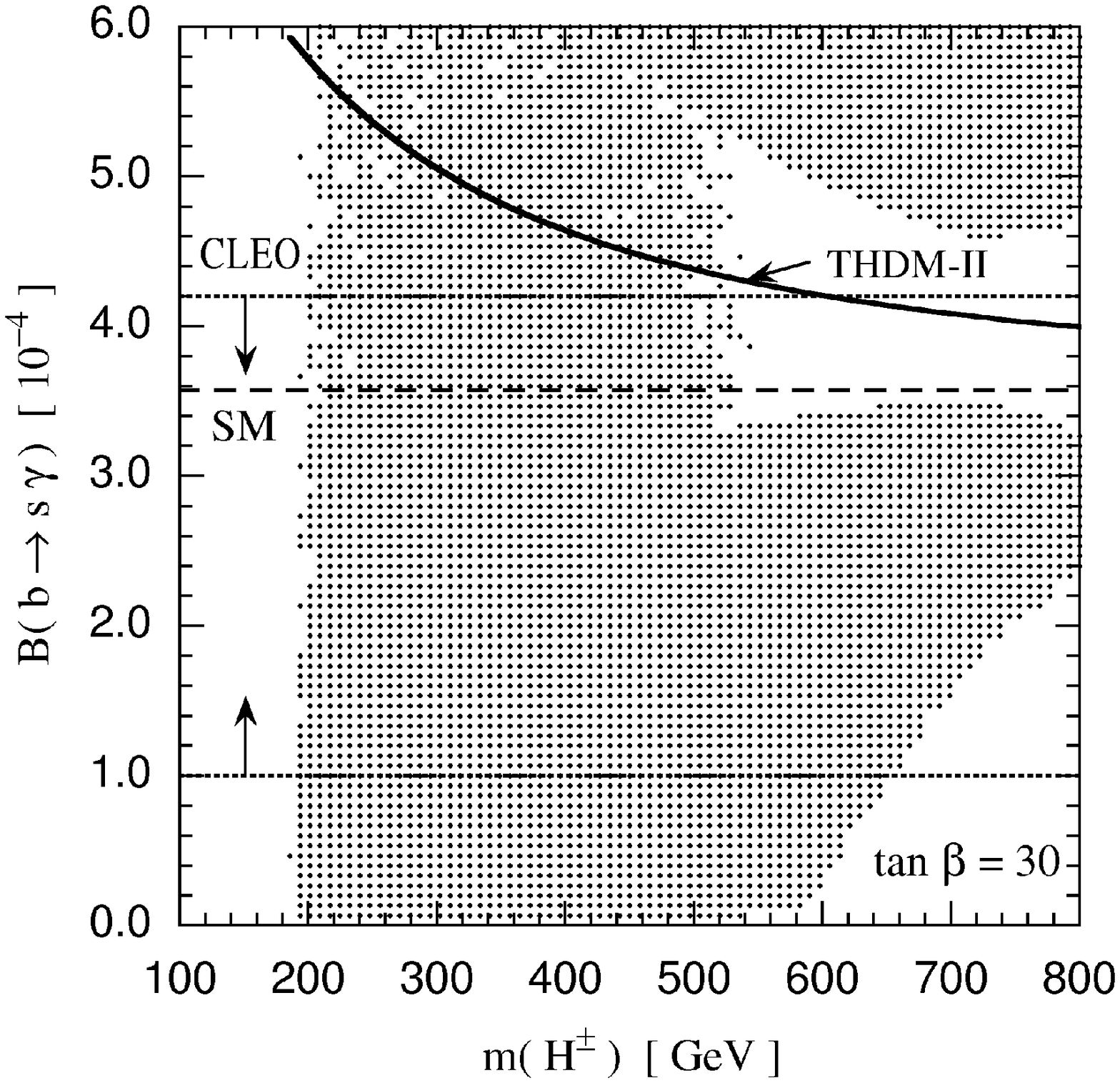}
}
\vfill
{\Large\bf Fig.~\ref{fig:bsg-cH}(b)}
\end{center}
\clearpage

~
\vfill
\begin{center}
\makebox[0cm]{
\def\epsfsize#1#2{\EPSSCALE#1}
\epsfbox{\EPSDIR 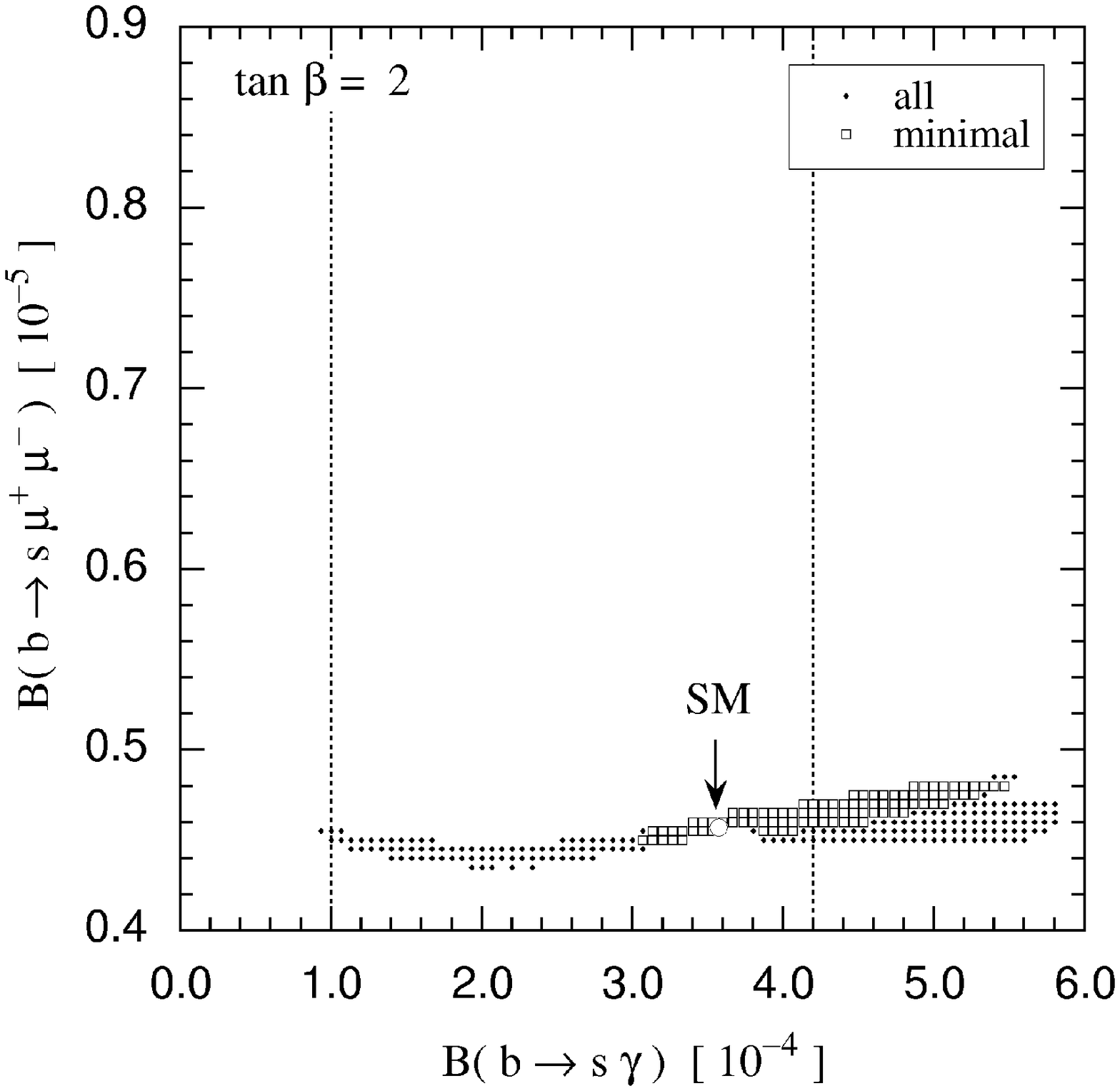}
}
\vfill
{\Large\bf Fig.~\ref{fig:bsll-bsg}(a)}
\end{center}
\clearpage

~
\vfill
\begin{center}
\makebox[0cm]{
\def\epsfsize#1#2{\EPSSCALE#1}
\epsfbox{\EPSDIR 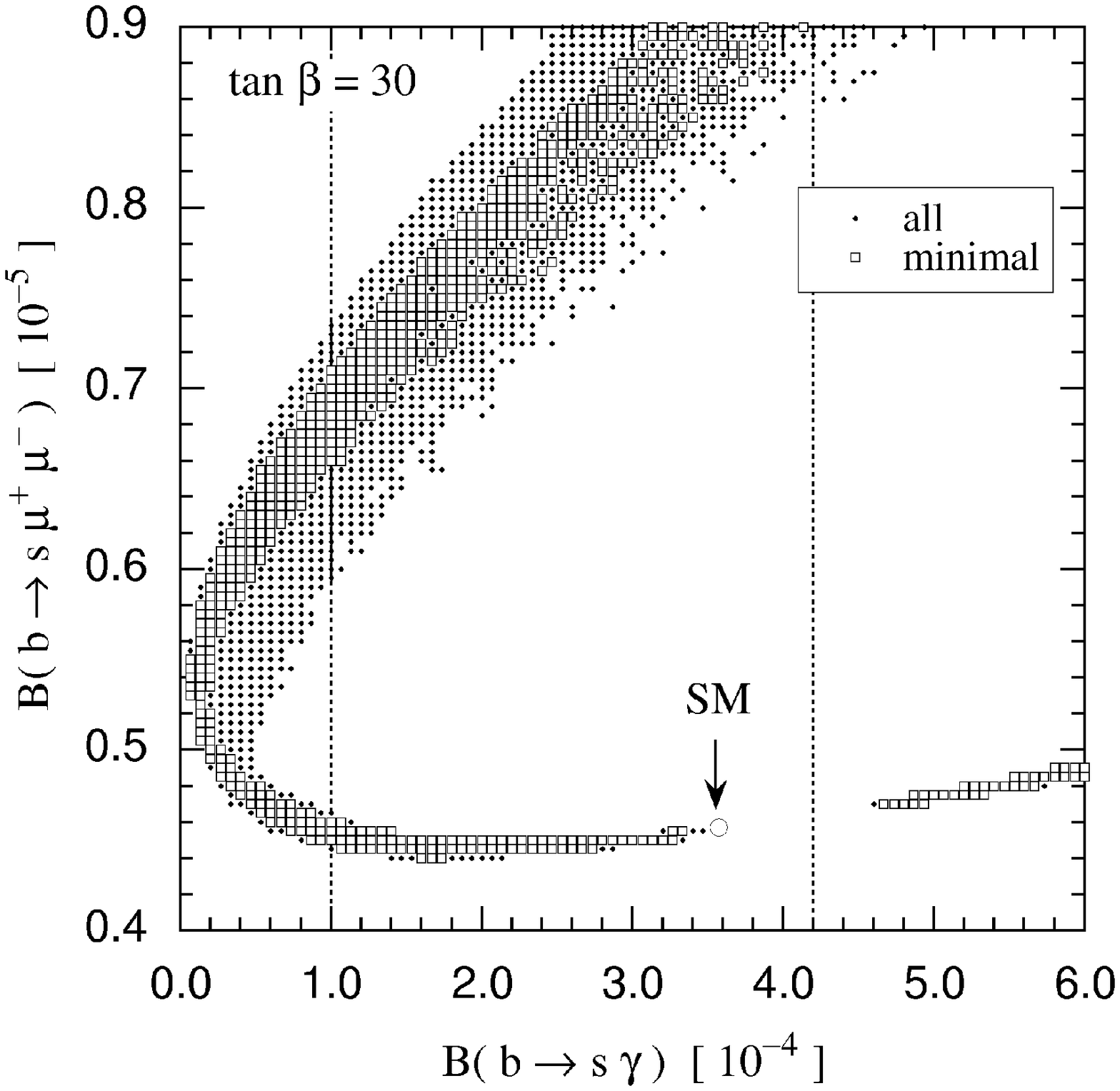}
}
\vfill
{\Large\bf Fig.~\ref{fig:bsll-bsg}(b)}
\end{center}
\clearpage

~
\vfill
\begin{center}
\makebox[0cm]{
\def\epsfsize#1#2{\EPSSCALE#1}
\epsfbox{\EPSDIR 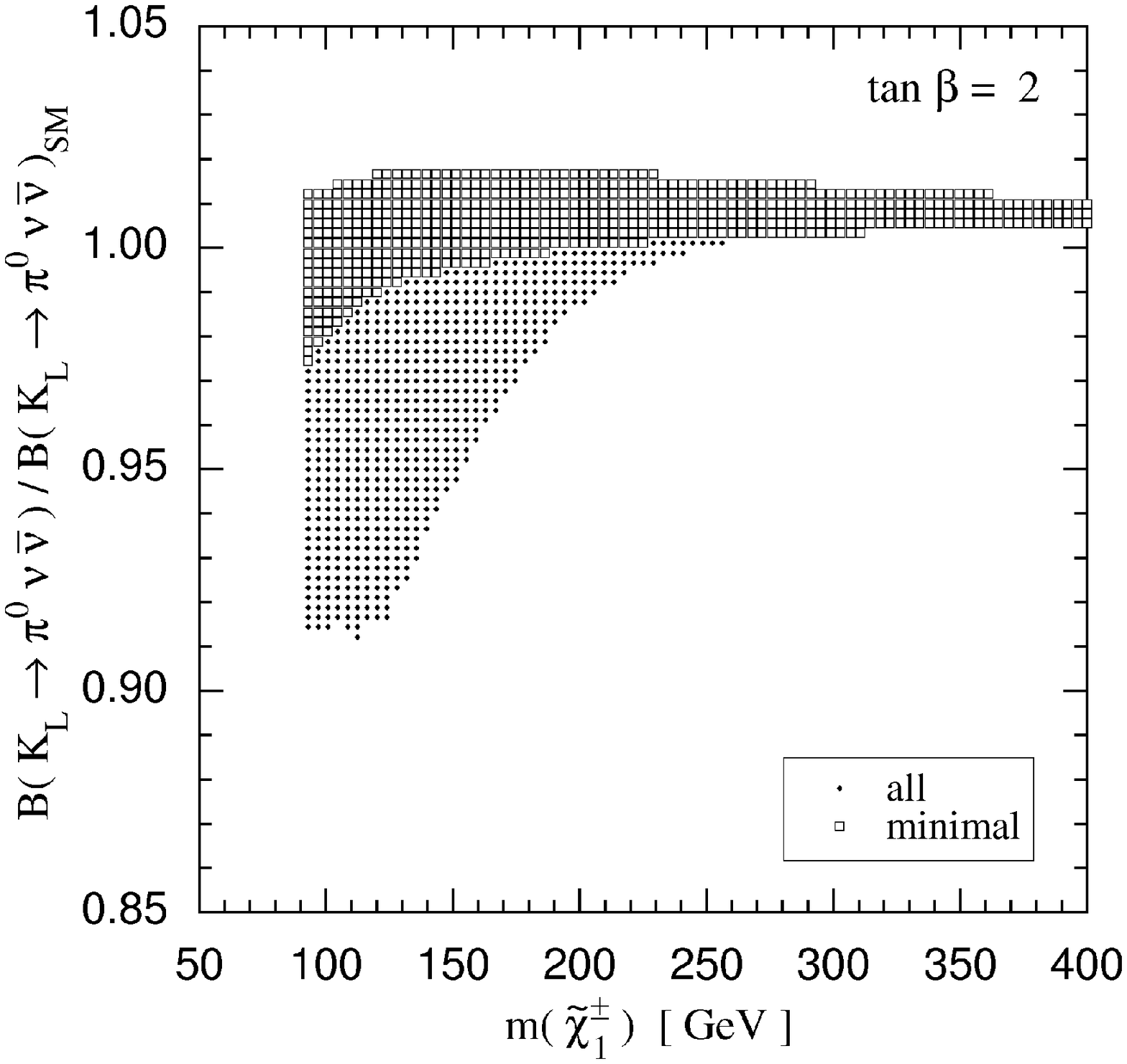}
}
\vfill
{\Large\bf Fig.~\ref{fig:kpnn-cno}(a)}
\end{center}
\clearpage

~
\vfill
\begin{center}
\makebox[0cm]{
\def\epsfsize#1#2{\EPSSCALE#1}
\epsfbox{\EPSDIR 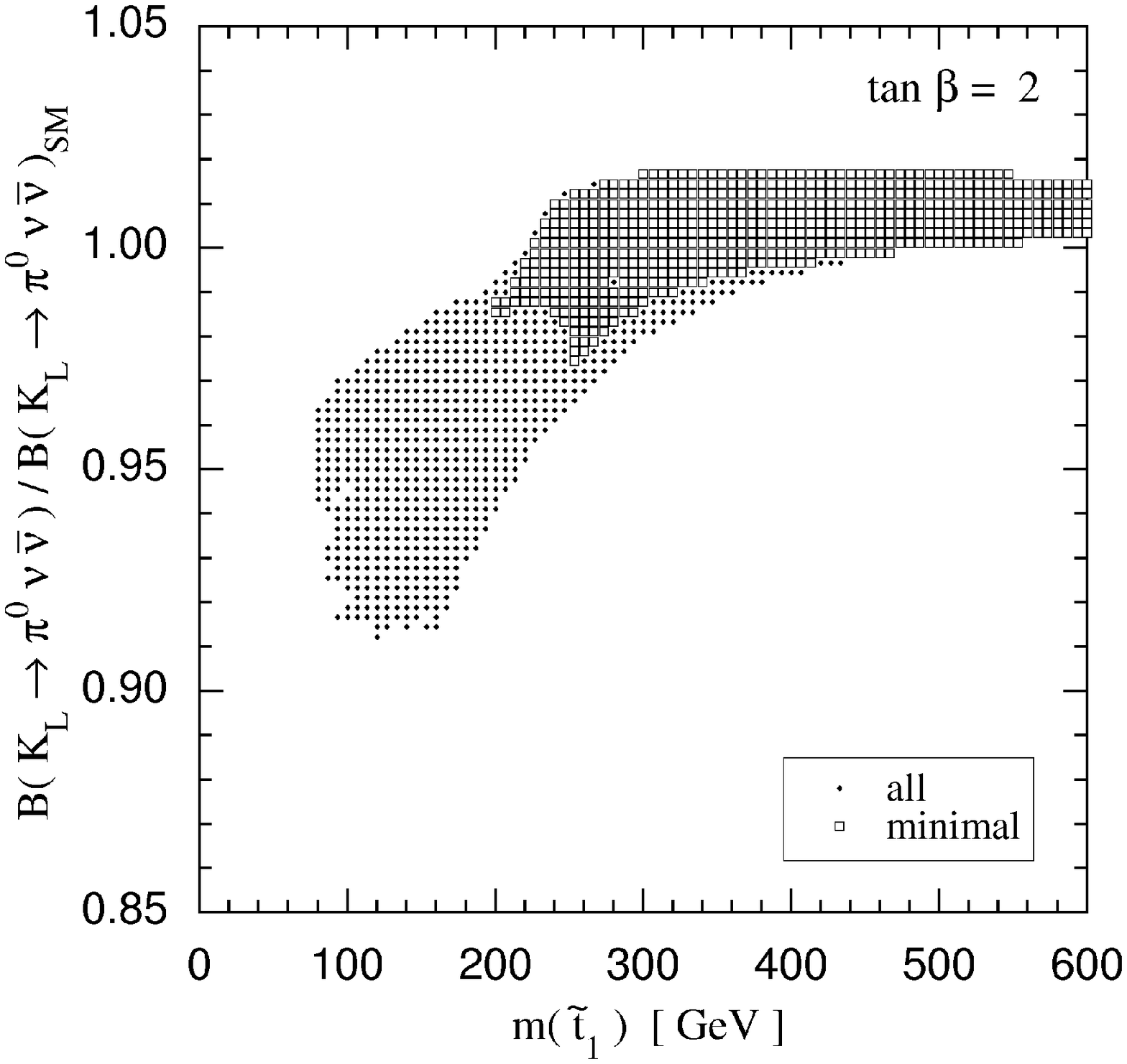}
}
\vfill
{\Large\bf Fig.~\ref{fig:kpnn-cno}(b)}
\end{center}
\clearpage

~
\vfill
\begin{center}
\makebox[0cm]{
\def\epsfsize#1#2{\EPSSCALE#1}
\epsfbox{\EPSDIR 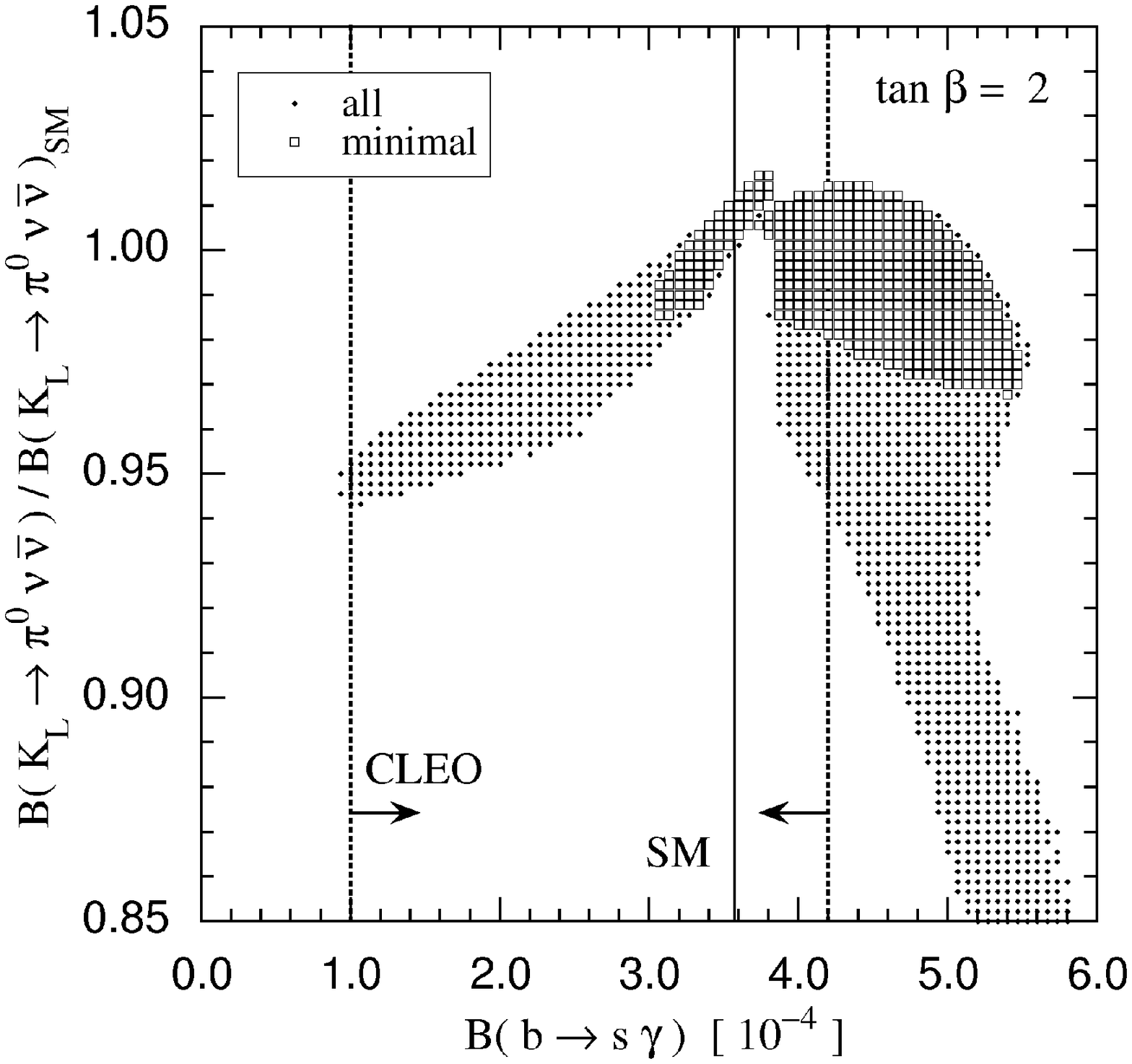}
}
\vfill
{\Large\bf Fig.~\ref{fig:kpnn-cno}(c)}
\end{center}
\clearpage

~
\vfill
\begin{center}
\makebox[0cm]{
\def\epsfsize#1#2{\EPSSCALE#1}
\epsfbox{\EPSDIR 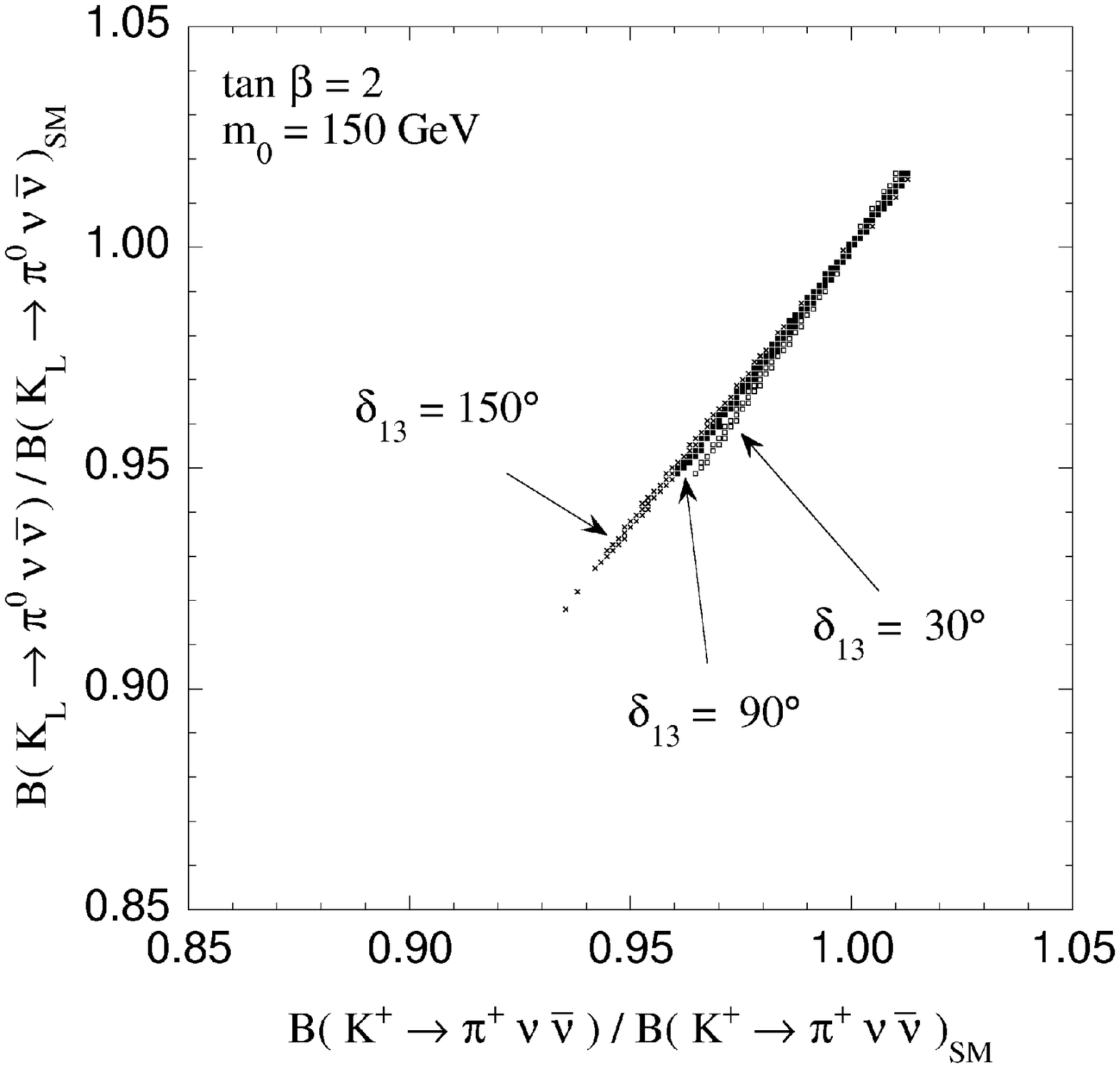}
}
\vfill
{\Large\bf Fig.~\ref{fig:klpnn-kppnn}}
\end{center}
\clearpage

~
\vfill
\begin{center}
\makebox[0cm]{
\def\epsfsize#1#2{\EPSSCALE#1}
\epsfbox{\EPSDIR 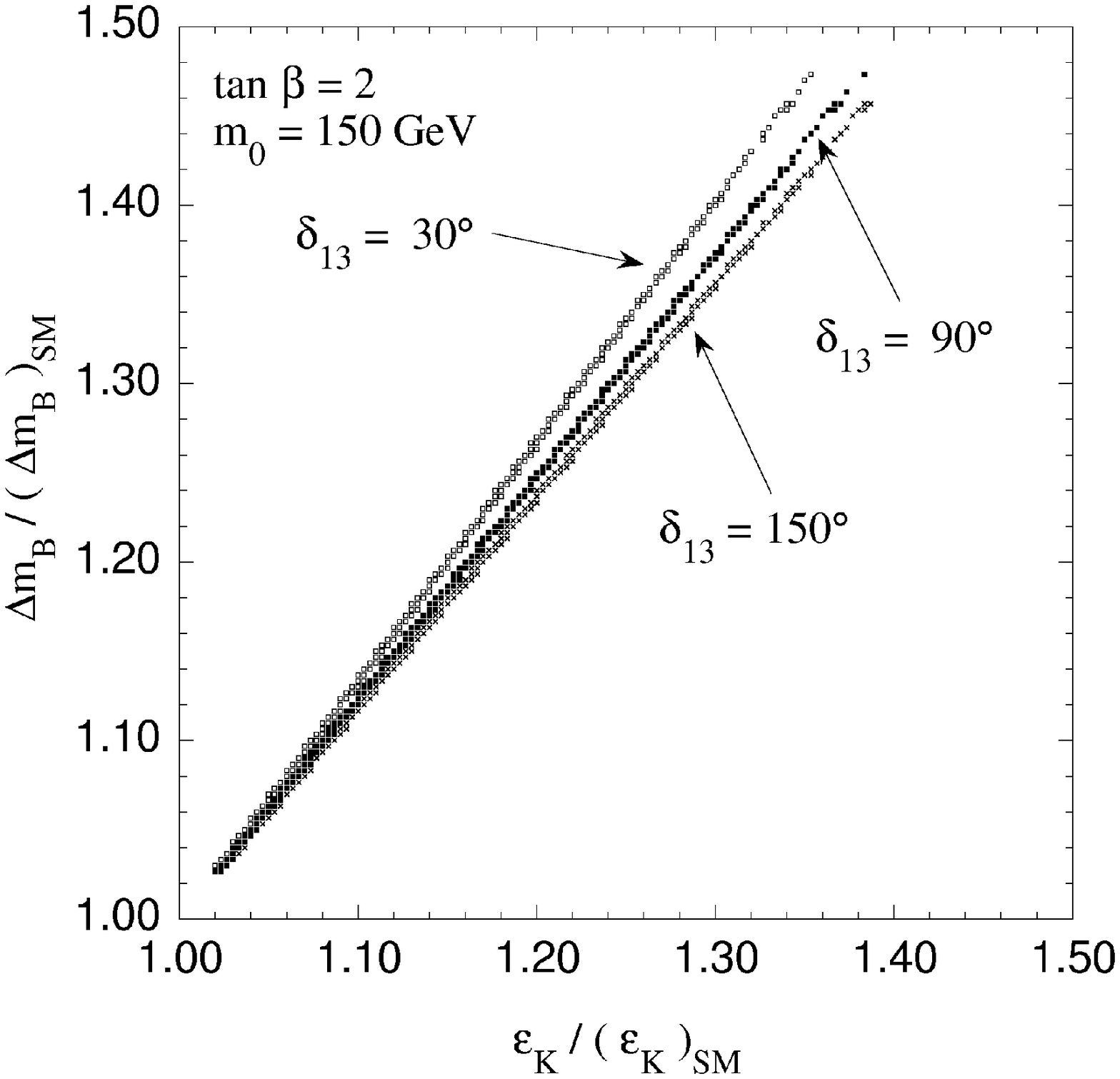}
}
\vfill
{\Large\bf Fig.~\ref{fig:ek-dmb}}
\end{center}
\clearpage

~
\vfill
\begin{center}
\makebox[0cm]{
\def\epsfsize#1#2{\EPSSCALE#1}
\epsfbox{\EPSDIR 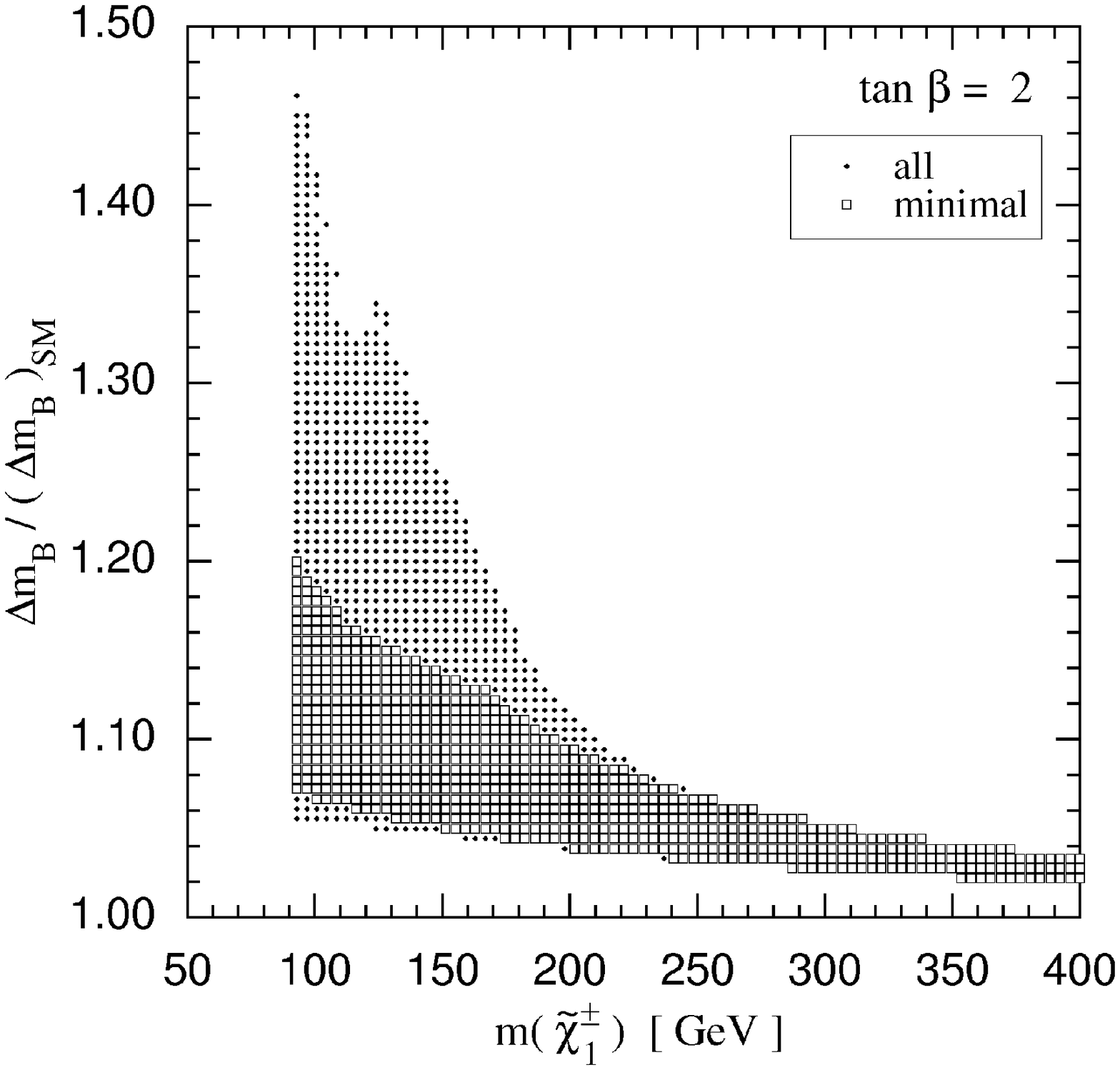}
}
\vfill
{\Large\bf Fig.~\ref{fig:dmb-cno}(a)}
\end{center}
\clearpage

~
\vfill
\begin{center}
\makebox[0cm]{
\def\epsfsize#1#2{\EPSSCALE#1}
\epsfbox{\EPSDIR 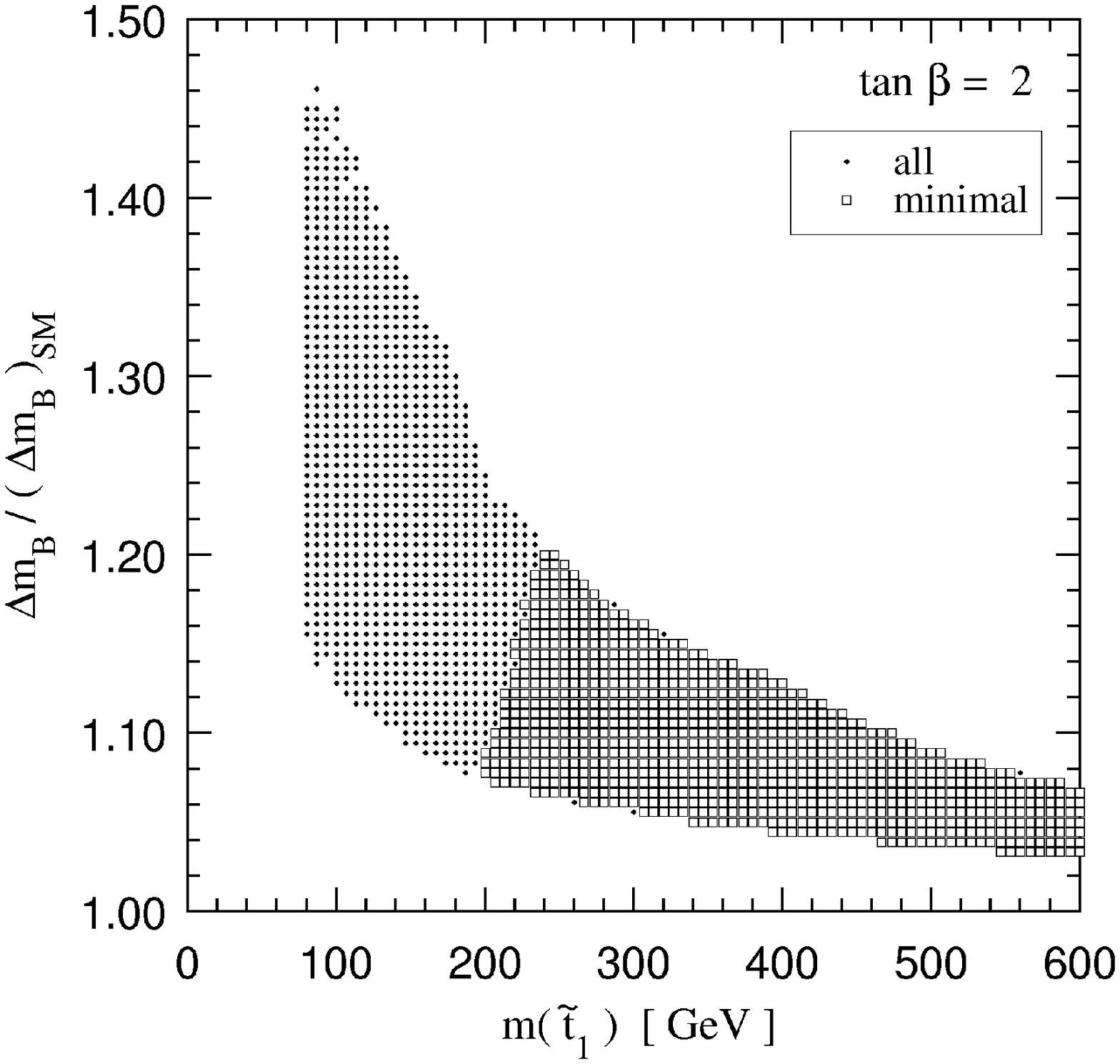}
}
\vfill
{\Large\bf Fig.~\ref{fig:dmb-cno}(b)}
\end{center}
\clearpage

~
\vfill
\begin{center}
\makebox[0cm]{
\def\epsfsize#1#2{\EPSSCALE#1}
\epsfbox{\EPSDIR 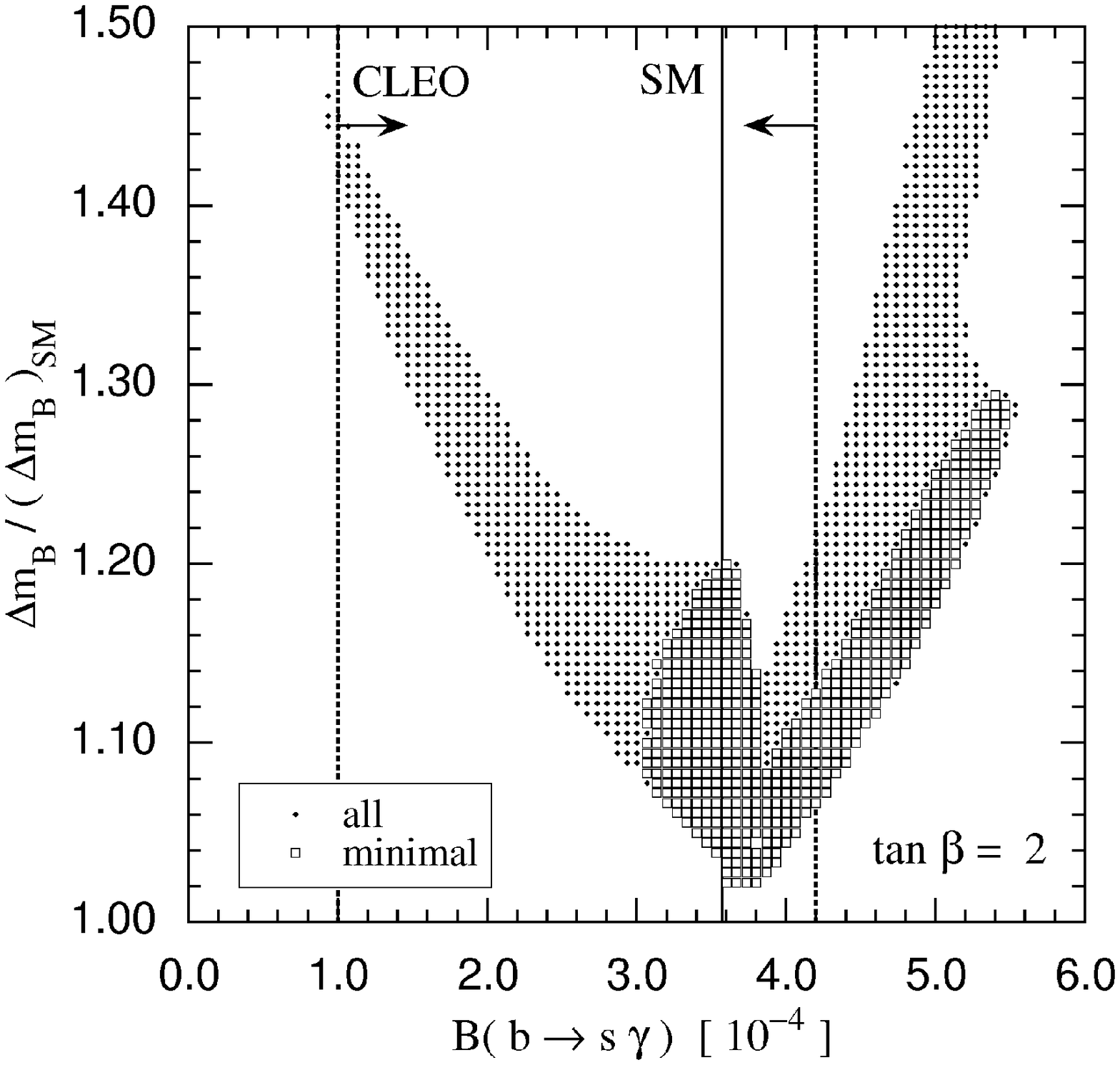}
}
\vfill
{\Large\bf Fig.~\ref{fig:dmb-cno}(c)}
\end{center}
\clearpage

~
\vfill
\begin{center}
\makebox[0cm]{
\def\epsfsize#1#2{\EPSSCALE#1}
\epsfbox{\EPSDIR 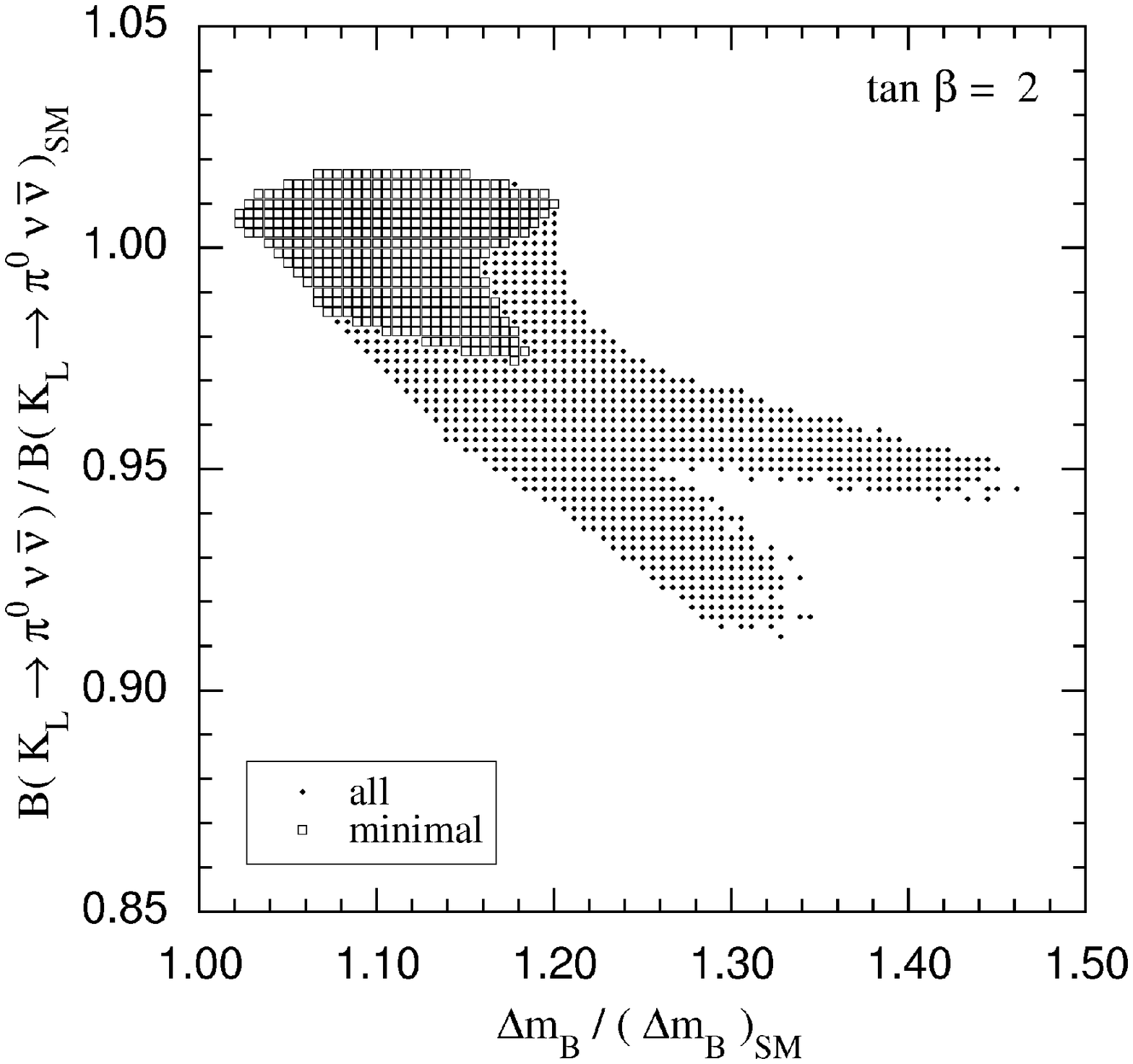}
}
\vfill
{\Large\bf Fig.~\ref{fig:kpnn-dmb}(a)}
\end{center}
\clearpage

~
\vfill
\begin{center}
\makebox[0cm]{
\def\epsfsize#1#2{\EPSSCALE#1}
\epsfbox{\EPSDIR 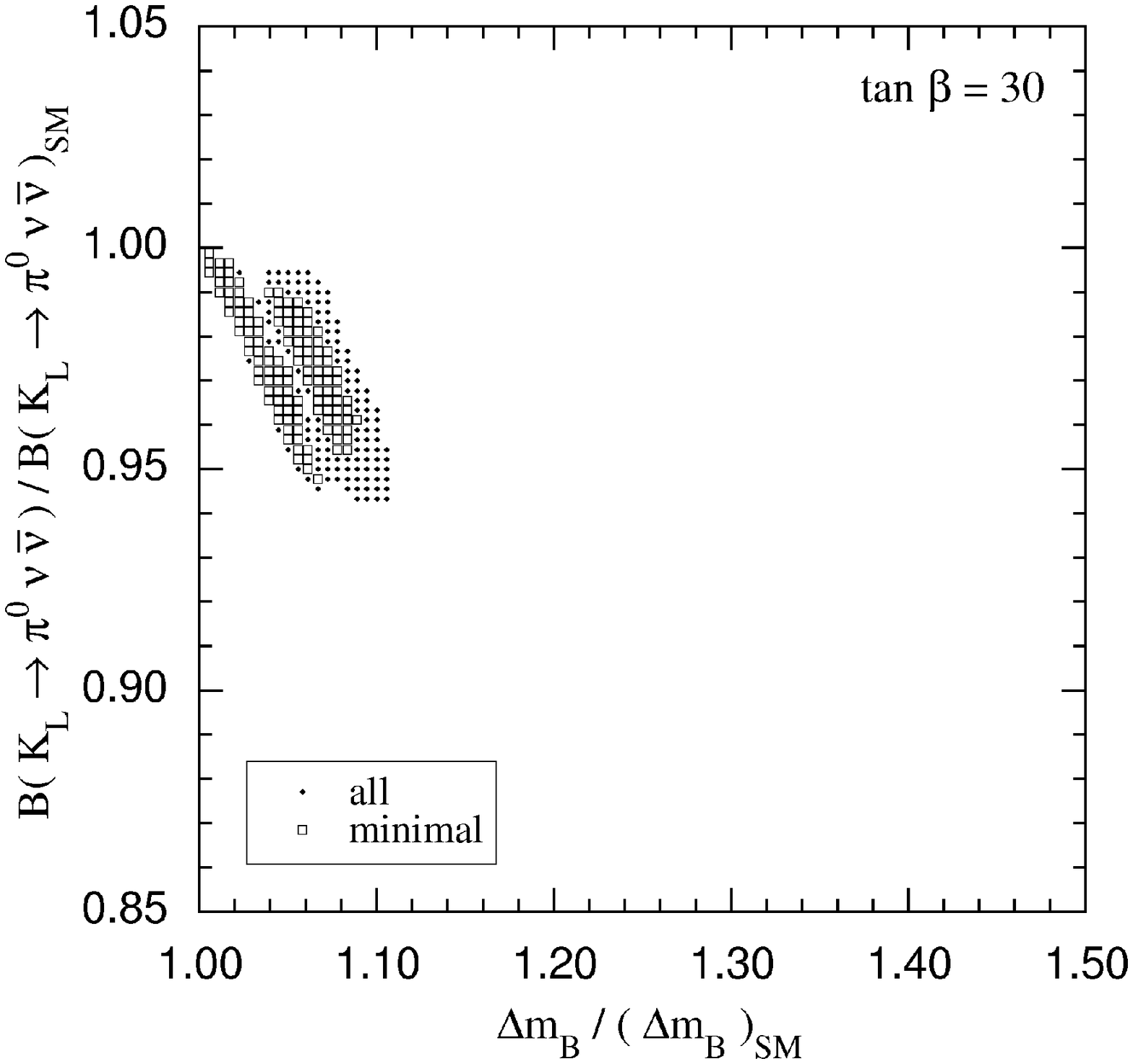}
}
\vfill
{\Large\bf Fig.~\ref{fig:kpnn-dmb}(b)}
\end{center}
\clearpage

\end{document}